\documentclass[namedreferences]{solarphysics}

\usepackage[optionalrh,solaromanenum]{spr-sola-addons} 

\usepackage{graphicx}        
\usepackage{color}           
\newcommand{\arcsec}{\hbox{$^{\prime\prime}$}}
\newcommand{\ion}[2]{#1\,{\sc #2}}

\newcommand{\farcs}{\mbox{\ensuremath{.\!\!^{\prime\prime}}}}%

\newcommand{\aap}{    {\it Astron. Astrophys.}}

\newcommand{\apj}{    {\it Astrophys. J.}}
\newcommand{\apjl}{   {\it Astrophys. J. Lett.}}

\newcommand{\solphys}{{\it Solar Phys.}}

\newcommand{\ssr}{    {\it Space Sci. Rev.}}
\chardef\us=`\_

\newcommand*{\ads}[1]{\href{\detokenize{#1}}{{\color{blue} ADS.}}}
\newcommand*{\doi}[1]{\href{http://dx.doi.org/\detokenize{#1}}{{\color{blue} DOI.}}}
\begin{document}

\begin{article}
\begin{opening}

\title{Two strong white-light solar flares in AR NOAA 12673 as potential clues for stellar superflares}

\author[addressref=aff1,email={paolo.romano@inaf.it}]
{\inits{P.}\fnm{P.}~\lnm{Romano}\orcid{0000-0001-7066-6674}}

\author[addressref=aff2,email={elmhamdi@ksu.edu.sa}]
{\inits{A.}\fnm{A.}~\lnm{Elmhamdi}\orcid{0000-0002-5391-4709}}

\author[addressref=aff2,email={askordi@ksu.edu.sa}]
{\inits{A.S.}\fnm{A.S.}~\lnm{Kordi}}

\address[id=aff1]{INAF - Osservatorio Astrofisico di Catania, Via S. Sofia 78, 95123, Catania, Italy}
\address[id=aff2]{Department of Physics and Astronomy, King Saud University, PO Box 2455, Riyadh 11451, Saudi Arabia}

\runningauthor{P. Romano et al. \textit{et al.}}

\runningtitle{3D null points in two WLFs.}

\begin{abstract}

Recently, two strong homologous white light flares of X-GOES class occurred on the Sun on Sept. 06, 2017, providing a rare exceptional opportunity to study the mechanisms responsible for the formation of the magnetic field configurations suitable for the manifestation of such yet enigmatic eruptive events and their effects in the lower layers of the solar atmosphere.

Using photospheric vector magnetograms, taken before the beginning of the two X-class events, as boundary conditions to reconstruct the non$-$linear coronal magnetic field configuration, we identified two related 3D null points located at low heights above the photosphere ({\it i.e.} in very low corona). These null points are most likely responsible for the triggering of the two strong X-GOES class flares. We deduced that their formation at such low altitudes may plausibly be ascribed to the peculiar photospheric horizontal motions of the main magnetic structures of the hosting Active Region NOAA 12673.

These events can be adopted as a hint for a possible interpretation of the activity of young G-type stars, recently reported by the $Kepler$ mission. We argued that a possible explanation of the acceleration of huge numbers of particles producing white light emission, during the Sept. 6 events as well as during white light flares in young Sun-like stars, might be attributed to the special accompanying conditions of the occurrence of magnetic reconnection at low altitudes of their atmospheres.

\end{abstract}


\end{opening}
\section{Introduction}

The most energetic flares occurring on the Sun are dynamical phenomena involving the whole solar atmosphere with important effects at photospheric level. During these events electron beams, characterized by particular high energy, reach the chromosphere which in turns, owing to its radiative heating, stimulates the photospheric emission (white light flares) (WLFs; \citet{Sve66}). The involved physical mechanism for the energy transportation from the upper chromosphere to the photosphere where the WL emission takes place is well known as ``back-warming effect" \citep{Mac89}. The duration of the optical continuum emission at photospheric level is found to vary, but usually associated to the impulsive phase of the corresponding flares. Worth noticing here that it is widely accepted by the scientific community that the white light emission is a common feature of all solar flares (see \citet{Mat03, Jes08, Hud06}), with the more energetic is the flare, the more detectable is the white light contribution. In fact, the energy required to power the white light emission can be similar to the total energy carried by an electron beam \citep{Met03}. However, the need of a better and deeper understanding of the physics behind the mechanisms leading to the acceleration of particles and hence the increase of the emission in the continuum of the solar spectra or in white light images remains a challenging topic and one of the highest priorities for the worldwide solar physics community investigation (see {\it e.g.} \citet{Mac89, Fan95, Din03, Rom18}).

Interestingly, the $Kepler$ mission has recently revealed the occurrence of extremely energetic flares, named ``superflares", on G-type stars. These superflares are characterized by energies several orders of magnitude greater than the observed strongest solar events, {\it i.e.} up to 10$^{38}$ erg. They have been observed in $Kepler$ broad optical passband from 400 to 900 nm and, for this reason, according to the standard solar terminology, they are also dubbed ``WLFs". Additionally, they seem to occur more often on cooler G-type stars ($T_{eff}$ = 5100$-$5600 K) with faster rotational periods (from few hours to several days) than on hotter stars ($T_{eff}$ = 5600$-$6000 K) characterized by slower rotational periods ($>$ 10 days) \citep{Mae12, Shi13}. \citet{Bal15} found that these WLF's associated energies appear to be strongly correlated with stellar luminosity and radius. However, only limited attention was devoted to understanding the mechanisms of their emission.

In spite of the diversity in physical parameters and energy budget between solar flares and superflares, we believe that the study of solar WLFs can provide clues for the study of the stronger events occurring on other stars. Therefore, in order to understand why stars younger and smaller than the Sun are able to produce eruptive events at such a very high energy domain, which cannot in principle occur on the Sun \citep{Aul13}, we need to scrutinize the characteristics and peculiarities of the observed strongest events recorded to date in the solar atmosphere. In fact, the study of the magnetic field configurations able to host the strongest events occurring on the Sun, as well as the involved precursor structures and topologies, can shed light even on the mechanisms at the base of superflares where a more efficient acceleration of particles seems to take place. Typical examples of strong solar events able to show clearly their effect in the low solar atmosphere are the sequence of X-class flares recorded from mid-October to early November 2003 (peaked on October 28), with a flare magnitude estimated as large as X45-class \citep{Tho04}, or the famous Bastille Day Flare (X5.7) occurred on July 14, 2000. In these cases, it has been observed that the bright ribbons exceeding the brightness of the surrounding photosphere by up to 20-30\% were located above or near the main sunspots. In particular, in the Bastille Day Flare, \citet{Gre08} found that the regions involved in the emission in different spectral ranges seemed to be associated with particle acceleration arising at low heights above the main sunspots. Various observations support the idea that their continuum emission at photospheric level may be correlated with low heights of the reconnection sites in the solar atmosphere \citep{Ni16}.

In order to provide further evidences and contributions in favor of this scenario and possibly probe the conditions for the magnetic reconnection to occur at low altitudes, we study, in the present work, two interesting homologous WLFs of X-class (X2.2 and X9.3) hosted by AR NOAA 12673. These events occurred on 2017 September 06, with about 3-hr time interval, between their peaks: the first flare peaked at 09:10 UT, while the second one at 12:02 UT (Figure 1). They offer a rare opportunity to explore two energetic events characterized by similar configuration of the involved magnetic systems but with different extent of their effects at photospheric level. The intriguing evolution of AR NOAA 12673 were subject of recent devoted efforts and a series of investigations \citep{Sun17, Yang17, Rom18, Ver18, Wan18, Yan18}. \citet{Sun17} for example highlighted the relevance of the unusual and significant emerging new magnetic flux in triggering the flares. Some aspects of the magnetic field topology and the strong transverse magnetic fields have been reported by \citet{Yang17}, \citet{Rom18} and \citet{Wan18}. The main peculiarity of these events were the strong horizontal displacement of part of the main sunspot of the active region (AR) that seems to have an important role in the storage of the magnetic free energy \citep{Rom18}.
Hence, these two homologous WLFs can be considered as interesting targets to understand the role of the sunspot motions in WLFs as well as to examine the mechanisms leading to the formation of the main magnetic reconnection site at low height above the photosphere. Moreover, our analysis of these events is also used to suggest a possible explanation of the strong activity in young G-type stars and their white light emission during superflares.

We studied the conditions that lead to the occurrence of these events by the dataset and the methods of analysis described in Section 2, focusing on determining the location of 3D null points, aiming to inspect and comprehend the related coronal magnetic conditions able to generate recurrent WLFs. We highlighted the main results in Section 3. Finally we discussed and interpreted our findings in Section 4.

\section{Data and analysis}

The AR NOAA 12673 has been observed to sojourn the solar disk on Sept. 2017. Among the 40 C, 20 M and 4 X-class flares unleashed by the AR, we analyzed the physical conditions for the occurrence of the strong X2.2 and X9.3 class flares occurred on Sept. 06 during a time interval of few hours (Figure 1). In particular, the most energetic event excited an unusual $sunquake$ phenomena. More details about the timing of the two events have been reported in \citet{Rom18}.

In this work, we focused on the coronal magnetic configuration at the base of such strong events developed in rapid succession in the same AR. In particular we highlight the various peculiarities of the two events, addressing the point of identifying the 3D null points as a key ingredient in understanding the magnetic topology associated to the WLF episode.  In order to reconstruct the 3D coronal magnetic field configuration we adopted the non-linear force-free (NLFF) extrapolation method proposed by \citet{He11} and based on the direct boundary integral equation (DBIE) formulation \citep{Yan06}. We used as boundary conditions for the NLFF field extrapolations the vector magnetograms taken by HMI \citep{Sch12} onboard the Solar Dynamic Observatory (SDO; \citet{Pes12}) in the \ion{Fe}{I} line at 617.3 nm  and pixel size of 0$\farcs$51. We utilized the data in the Space-weather Active Region Patches (SHARPs) version \citep{Hoe14} taken on Sept. 06 at 00:00 UT, 08:48 UT and 11:48 UT, exploiting hence the reconstruction of the magnetic field configuration in different phases ({\it i.e.} before, after and between the peaks). The input 2D photospheric vector magnetograms for the extrapolations process were 70 $\times$ 70 pixels (rebinned data) with a field of view of 176 $\arcsec$ $\times$ 176 $\arcsec$, corresponding to a resolution of 2$\farcs$51/pixel. The output 3D coronal magnetic fields were 70 $\times$ 70 $\times$ 40 grids with the photospheric magnetogram being retained in the output 3D data as the bottom layer (layer 0) of the data cube.

We applied the method outlined in \citet{Bev06} to the output NLFF field data cubes in order to determine the coronal null points, {\it i.e.} the locations where the magnetic field vanishes and the reconnection can take place \citep{Pri00}. This method allows indeed to find 3D null points in two steps. First, it applies a spine graph analysis which reduces the complicated topological footprint to a graph theory problem and infers the existence of almost all of the separators, {\it i.e.} the field line connecting two null points which comes from the intersection of two separatrix surfaces of different magnetic systems. Second, it employs a simulated annealing method: in analogy to annealing in crystals, in which the lowest-energy state is found over large length scales by making small random perturbations to its current state, the method permits to assess which nulls are connected by separators. Finally, among the nulls found by the algorithm we selected only those located above the boundary conditions of the extrapolations.

Taking into account the noticeable horizontal displacement of the negative polarity of the main sunspot of the AR, we also used vector magnetograms taken with a time interval of 24 min to derive the horizontal velocity fields by means of the DAVE4VM method. As described in \citet{Sch08}, the velocity was modeled by a variational principle to minimize deviations in the magnitude of the magnetic induction equation constrained by a 3D affine velocity profile, which depends linearly on coordinates, within a windowed subregion of the magnetogram sequence. In this case we used a full width at half maximum of the apodization window of 11 pixels.

Furthermore, and to highlight the effect of the photospheric horizontal velocities on the overlying magnetic systems, we follow the approach of \citet{Fal02} and \citet{Jia14} to measure the shear angle, {\it i.e.} the angle between the observed horizontal field and the horizontal field derived through a potential
field extrapolation. We adopted the horizontal shear angle equation defined as \citep{Gos10}:
\begin{equation}
\theta=arcos\frac{\bf B_{h}^{obs} \cdot B_{h}^{pot}}{\left|B_{h}^{obs}\right|\left|B_{h}^{pot}\right|},
\label{Eq1}
\end{equation}
where $B^{obs}_{h}$ and $B^{pot}_{h}$ are the horizontal components of the observed magnetic field and the potential magnetic field such that $B^{pot}_{r}$=$B^{}_{r}$, respectively.

We also quantified the difference between the inclination angle of the observed field $\gamma^{obs}$ and that of the potential field $\gamma^{pot}$:
\begin{equation}
\Delta\gamma=\gamma^{obs}-\gamma^{pot},
\label{Eq2}
\end{equation}
named as ``dip angle", which can be considered as a measure of the vertical shear. In fact, the closer the field approaches the potential field configuration, the smaller is the value of the inclination difference \citep{Gos10}.

In addition, RHESSI data are used to reconstruct the signal for 12$-$25 keV and 25$-$50 KeV bands around the peaks of the two WLFs. We obtained a spatial resolution of 4 arcsec by means of the use of 4 detectors, that were operational during the observation period of these flares. We applied the CLEAN method \citep{Hur02} with an integration interval of 4 s.


\section{Results}

In the continuum images observed by SDO$/$HMI, AR NOAA 12673 appears to be constituted by a main delta spot ($S1$), surrounded by some smaller spots ($S2$-$S5$) in the Northern and Southern parts of the AR (left panel of Figure 2). In the delta spot we distinguish four umbrae. Three of them, indicated by $U1$, $U2$ and $U3$ in the middle of left panel of Figure 2, have positive polarity, while the eastern one, marked by $U4$, has negative polarity and appears on Sept. 06 at 00:10 UT as divided by a thin light bridge.

Strong shear motions have been observed in $S1$ before the two homologous WLFs as well as during the time interval between them. In particular, $U4$ showed a displacement of tens of arcsec in the northern direction, with an average velocity of 0.4 km s$^{-1}$ for several hours, hence producing some variations of the coronal magnetic field configuration yielding to favorable conditions for the occurrence of the strong events with effects in the low layers of the solar atmosphere (see \citet{Rom18} for further details). Two snapshots of the main delta spot before the beginning of the two flares, with the velocity fields computed adopting the Differential Affine Velocity Estimator method for vector magnetograms \citep[DAVE4VM;][]{Sch08}, are reported in Figure 3. Interestingly, the northward displacement of the umbra $U4$ was in addition accompanied by the southward motions of the magnetic features, forming the remaining part of the sunspot ($U1$)(compare Figures. 2 and 3). In particular, the northern portion of the negative polarity, which seems to divide the positive polarity in two parts, shows velocities up to 0.7 km s$^{-1}$ at 11:36 UT, {\it i.e.} few minutes before the beginning of the second WLF (see the right panel of Figure 3).

This particular behavior of the photospheric features is also confirmed through the topology of the transverse magnetic field. In the top panel of Figure 4 the transverse field vectors are overplotted on a B$_z$ map background (in Cylindrical Equal Area ($CEA$) coordinates), while the bottom panels are the zoomed views of the region of interest ({\it i.e.} enclosed by a box in top panel), before the first peak (at 08:36 UT), between the two peaks (at 10:00 UT) and hours after the second peak (at 17:00 UT). The strong shear topology along the PIL, as well as the evident counterclockwise rotation of the north-eastern component of the negative polarity are evidently highlighted by  the structure and temporal evolution of the horizontal fields. Both the transverse and velocity fields maps clearly indicate the AR being theater of strong and long-continued photospheric motions accompanied with a substantial shear along the main PIL. It is noteworthy that these motions persist for about 24 hours until the end of the second flare, suggesting the shear motions being an important mechanism capable of supplying new energy into the magnetic systems of the AR involved by the flares.

The main peculiar characteristics of the two flares hosted by AR NOAA 12673 on Sept. 06, 2017, are their homology and emission in the continuum. Although the energy delivered by these flares (X2.2 and X9.7 GOES class) is not exceptional, these two reported characteristics undoubtedly make them unique targets for the interpretation of the different amount of energy involved in homologous events and its deposition at photospheric level.

The enhancement of the emission at photospheric level of the two events is shown in the continuum by HMI during the flare peaks. Exploiting the HMI resolution we detected small kernels of emission over the main spot during the X2.2 class flare (left panel of Figure 5). The area of these kernels reaches a total maximum extension of about 2.4 $\times$ 10$^{6}$ km$^{2}$. Obviously, the photospheric emission of the second flare is more intense and exhibits a clear ribbon shape, including the same regions previously involved by the kernels of the first flare (right panel of Figure 5). The area covered by the ribbons in the continuum filtergram at 11:59 UT reaches a maximum extension of about 9.2 $\times$ 10$^{7}$ km$^{2}$. The ribbons are located at both sides of the PIL: the eastern one runs along the umbra $U4$, while the western one, almost parallel to the other ribbon, reaches the south-western edge of the main spot.

The homology of the two WLFs is demonstrated not only in the similar morphology of the chromospheric ribbons, but also in the brightening of the coronal magnetic systems of the AR observed during the main phases of both events.

In the top left panel of Figure 6 we display the ribbons observed by INAF-Catania Astrophysical Observatory on Sept. 06, 2011, at 09:31 UT, where three main kernels, indicated by $R1-R3$, are clearly distinguishable. Only the kernel $R1$ had a correspondent kernel at photospheric level (compare the left panel of Figure 5 and the top left panel of Figure 6). In the top left panel of Figure 6 we also marked the location of the flux peak at 12$-$25 keV taken by RHESSI at 9:10 UT, {\it i.e.} around the peak of the first WLF observed by GOES at 1$-$8 \AA. It is noteworthy that the hard X-ray source is located above the ribbon $R1$ in the center of the delta spot, {\it i.e.} where the small photospheric kernels have been observed by HMI. The H$\alpha$ ribbons of the X9.7 class flare have almost the same location and shape like those in the first flare (top right panel of Figure 6). In addition, we note a newly appearing smaller ribbon, indicated as $R4$. Its location corresponds to the eastern ribbon in the continuum filtergrams (compare the right panel of Figure 5 and the top right panel of Figure 6). The flux peak at 12:09 UT, observed by RHESSI at 12$-$25 keV, appears again over the ribbon $R1$, although this time it is closer to the PIL and to the new ribbon $R4$.

Thinner homologous ribbons are also observed at 1600 \AA{} by AIA (bottom panels of Figure 6). In this spectral line, emitted by the C \small{IV}, we observe the contribution of the upper photosphere and the transition region. The main difference between the two flares at this wavelength is that a new ribbon is visible during the second flare, as already noted in the continuum and in the H$\alpha$ line.

The observation of the AR at coronal level confirms the homology of the two flares too. In fact, the images taken at 211 \AA{} by AIA before the beginning of the WLFs, reported in Figure 7, show the brightening of the same bundles of loops at 8:39 UT and 11:45 UT. An interesting feature, mainly visible at 8:39 UT, is an helical structure crossing the AR from [-230$\arcsec$,500$\arcsec$] to [-300$\arcsec$,540$\arcsec$]. The same feature appears at 11:45 UT, although in this case it is covered by some bright and straight loops located along the north$-$south direction.

The presence of these stressed features is also confirmed by the maps of the evaluated shear and dip angles reported in Figure 8. A shear angle greater than 90$^{o}$ is visible in a wide portion of the main sunspot along the PIL. In particular, we note very high values of the shear angle, exceeding 120$^{o}$, along the umbra of negative polarity and in the northern portion of the positive polarity, both at 8:46 UT and at 11:46 UT (top left and top right panels of Figure 8, respectively). An interesting alternation of positive and negative dip angle values crossing the PIL from east to west documents also the importance of the vertical shear of the magnetic field along the whole PIL, probably as a consequence of the observed horizontal velocity fields (see the bottom panels of Figure 8).

To further investigate and determine the magnetic topology associated to the two WLFs as well as to understand how the strong horizontal motions are able to produce appropriate conditions for the occurrence of the magnetic reconnection at the base of these events, we performed three non-linear force-free extrapolations adopting as boundary conditions the vector magnetograms taken by the SDO/HMI instrument on Sept 06 at 00:00 UT, 08:48 UT and 11:48 UT, {\it i.e.} several hours before the two WLFs, few minutes before the beginning of the X2.2 flare and few minutes before the beginning of the X9.3 flare, respectively. These extrapolations are reported in Figure 9. Initially, in the area involved by the WLFs we distinguish three main arcades corresponding to different magnetic domains and to different connectivity of the photospheric magnetic structures ($``(a)"$ and $``(b)"$ panels of Figure 9). The negative umbra of the delta spot $U4$ seems to be connected to the positive polarity located in the southern part of the AR (blue lines) and corresponding to the spot $S5$ of Figure 2. Instead, the main positive umbrae of the delta spot ($U1$ and $U2$ in Figure 2) are linked to the negative pores $S3-S4$ in the northern part of the AR (yellow lines). A third arcade connects the positive polarity at north of the negative umbra ($U3$ in Figure 2) with the negative spot $S2$ at the northern edge of the main sunspot (red lines). About 8 hours later the magnetic topology of the corona appears to be clearly reconfigured (see panels $``(c)"$ and $``(d)"$ of Figure 9). A 3D null point is actually formed over the polarity inversion line between $U4$ and $U3$. Our computations indicate it to be located relatively low in the solar atmosphere, {\it i.e.} at about 5000 km above the boundary level. It is noteworthy that the projected position of this null point over the solar disk coincides with the contours indicating 70\% and 90\% of the flux peak at 12$-$25 keV taken by RHESSI. Moreover, we note that the ribbon locations observed in the chromosphere correspond to the footpoints of the magnetic flux systems determined from the extrapolation, namely: the ribbon $R1$ is located at the base of the northern footpoints of the field lines indicated by the blue lines, the ribbon $R2$ corresponds to the northern footpoints of the red lines while the ribbon $R3$ to the northern footpoints of the yellow lines. In principle, a reconnection occurring in that 3D null point can be responsible of the particle acceleration along the traced field lines and hence can explain the observed ribbons of the X2.2 class flare.

The third extrapolation shown in the panels $``(e)"$ and $``(f)"$ of Figure 9 is characterized by some small differences in comparison with the previous one. We recognize the same magnetic flux systems of the extrapolation obtained from the vector magnetograms taken before the first WLF, however, in this case, a new 3D null point appears near the position of the previous one and lower in the involved computational volume. Its position corresponds to about 3000 km above the photosphere, and is located over the small flux rope displayed by the red field lines, that connects the northern edge of the negative umbra ($U4$) with the contiguous positive polarity ($U3$). We remark that the involvement of this flux rope in the second events is confirmed by the new appearing ribbon $R4$ (upper right panel of Figure 6), which was not observed during the previous flare and which was located in an area corresponding to the eastern footpoints of the flux rope.


\section{Discussion and conclusions}

The homologous WLFs occurred on Sept. 06, 2017, hosted by AR NOAA 12637 allowed us to consider and emphasize some interesting aspects concerning the sources of the continuum emission observed at photospheric level and to provide some clue hints to interpret the analogous WLFs recently observed in young Sun-like stars by the $Kepler$ mission.

One of the main interesting aspects of the studied events is their homology, indubitably inferred by means of multiwavelength observations analysis as well as through force-free extrapolations performed using as boundary conditions the vector magnetograms taken by HMI in the \ion{Fe}\small{I} line at 617.3 nm.

The AIA images taken at 211 \AA{}, corresponding to the high corona at two million K, showed the involvement of the same magnetic systems during both flares. The brightest regions observed during the corresponding preflare phases, {\it i.e.} before that the flare emissions saturated the detector, highlighted indeed similar structures despite of both the three hours time interval between the two flare peaks and the reported significant displacement of the main photospheric magnetic structures. Moreover, the similar shapes of the ribbons of both flares at 1600 \AA{} and in the core of the H$\alpha$ line support the idea that the particle acceleration occurred along the same magnetic field lines reaching the lower layers of the solar atmosphere in the same regions. Therefore, we argue that the different emission at photospheric level observed during the peak of the two flares was most likely related to the different amount of free energy release. In effect, although the area covered by the continuum emission of the two events differed by one order of magnitude, the photospheric ribbons of the X9.7 class flare included the same regions previously covered by the kernels of the X2.2 class flare. This result suggests that the extent of the photospheric area affected by the chromospheric radiative heating may depend by the total energy of the electrons accelerated along the same field lines. It is also noteworthy that, contrary to the TiO emissions observed by \citet{Yur17}, in these flares the photospheric emissions appear co-spatial with chromospheric flare ribbons implying that both effects originate in the same chromospheric/photospheric volume.

A central key ingredient for the interpretation of the different amount of energy delivered by the two WLFs has been investigated through the analysis of NLFF extrapolations. Before the beginning of both flares a 3D null point in the low corona has been identified. Interestingly, many theoretical studies point out that 3D null points may be sites for magnetic reconnection \citep{Pri09, Gal11}. Moreover, their role in accelerating particles is supported by several simulations (see \citet{Bau13} and references therein). For this reason we think that the presence of these 3D null points, located exactly above the northern portion of the negative polarity of the main sunspot, where the higher values of horizontal velocity field have been observed, may have a crucial role in the flare trigger and in the acceleration of the electrons reaching the lower layers of the solar atmosphere. This is confirmed not only by the precise correspondence between the ribbons locations and the footpoints of the extrapolated magnetic systems, but more interestingly by the emission registered by RHESSI in the HXR at 12$-$25 keV corresponding to the location of the 3D null points. In fact, low-energy HXR sources have been interpreted in some previous studies as highly related to the outflow in vicinity of the reconnection site ({\it e.g.}, \citet{Sui03}).

We remark that, despite of the similar magnetic topology obtained by the NLFF extrapolations from the vector magnetograms taken few minutes before each flare, a different height of about 2000 km between the two 3D null points has been found. Taking into account that \citet{Ni16} simulations demonstrate that the continuum emission at photospheric level may be correlated with a low height of the reconnection site in the solar atmosphere, we think that in our homologous WLFs the different flare energy and the corresponding photospheric emission may mainly be attributed to the variation of the null point height: the lower the 3D null point, the stronger the flare.

There are three possible explanations of the variation of the null point height in time: (i)- the emergence of new magnetic flux which slightly changes the magnetic field topology of the AR in the low atmosphere, (ii)- the photospheric horizontal motions which stretches and lowers the magnetic field lines and (iii)- a combination of the previously cited two mechanisms. In our case, the magnetic field observations at photospheric level do not show the emergence of new polarity during the short time interval between the peaks of the two flares (only 3 hours), instead the northward motion of the negative polarity of the main spot reaches its maximum velocity (up to 0.7 km s$^{-1}$) at 11:36 UT (right panel of Figure 3). Consequently, we think that the stretching of the magnetic field lines due to the shear motions, observed for several hours before the flares and during the time interval between them, can be the main source of the lower height of the 3D null point, suitable for the trigger of the X9.7 class flare. In this regard, the maps of the shear and dip angle highly confirm the presence of considerable magnetic stresses along the PIL and in the regions corresponding to the strong brightening at chromospheric and coronal level. Therefore, and through our observations and analysis it seems that energetic flares occurring on the Sun, with evident effects in the continuum of the spectra, are linked to magnetic reconnection in the lower layers of the solar atmosphere \citep{Din99} and that this may happen when the magnetic systems are stretched by substantial and strong shear motions. Certainly, we cannot exclude that also the helical structure, visible at 211 \AA (Figure 7) and the corresponding green and grey arcades in the extrapolations (Figure 9), play an important role in the evolution of these events, however this is not in contrast with the above described scenario.

In this context, we believe that the shear motions observed along the PIL of the main sunspots, together with the high intensity of the involved magnetic flux systems, may be adopted as a reference for the study of flares occurring on other stars and characterized by higher orders of magnitude. Of course, two quite common X-class flares can not be assumed as a paradigm for the interpretation of all flares, but their homology and their emission in the continuum allow us to consider them as an interesting hint for a better comprehension of the possible triggering mechanisms of energetic flares/WLFs in other stars. The main characteristics of cooler G$-$type stars, where the $Kepler$ mission detected many superflares, recall the main ingredients at the base of the events subject of the present investigation, {\it i.e.}, high strength of the magnetic field and strong horizontal velocities. These G$-$type stars are characterized by an average magnetic field strength higher at least by an order of magnitude than that of the Sun at its maximum of activity \citep{Kat15}.  \citet{Not13} showed that many superflares probably occur in large starspots (10 times larger than the largest sunspot) and that the energy of the superflares can be explained by the magnetic energy stored around such starspots \citep{Shi13}. Moreover, stars with relatively slower rotation rates can still produce superflares that are as energetic as those of more rapidly rotating stars, but their average flare frequency is lower \citep{Noy84}. This is due to the fact that the fast rotation drives the emergence of strong magnetic fields.
However, we think that the height of the reconnection site may play also an important role in the amount of released energy. For this reason in the next future, we plan to analyze a wide sample of strong solar events, in order to effectively delineate a possible correlation between the height of 3D null points and the flare energy.

\begin{acks}
The HMI and AIA data have been used courtesy of NASA/SDO and the AIA and HMI science teams.
\end{acks}\\

{\bf Disclosure of Potential Conflicts of Interest} The authors declare that they have no conflicts of interest.

\begin{figure*}
 \hspace{0.5cm}\includegraphics[trim=0 0 0 0, clip, scale=0.9]{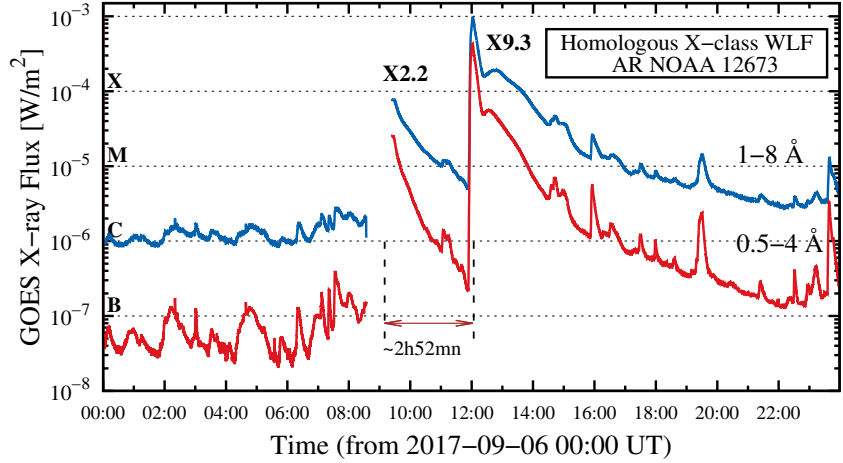}\\
  \caption{GOES X-ray light curves (blue curve: 1$-$8 \AA, red curve: 0.5$-$4 \AA ) during the two strong X-class flares of Sep. 09, 2017, hosted by AR NOAA 12673.\label{Fig1}}
\end{figure*}

\begin{figure*}
  \includegraphics[trim=5 100 170 300, clip, scale=0.4]{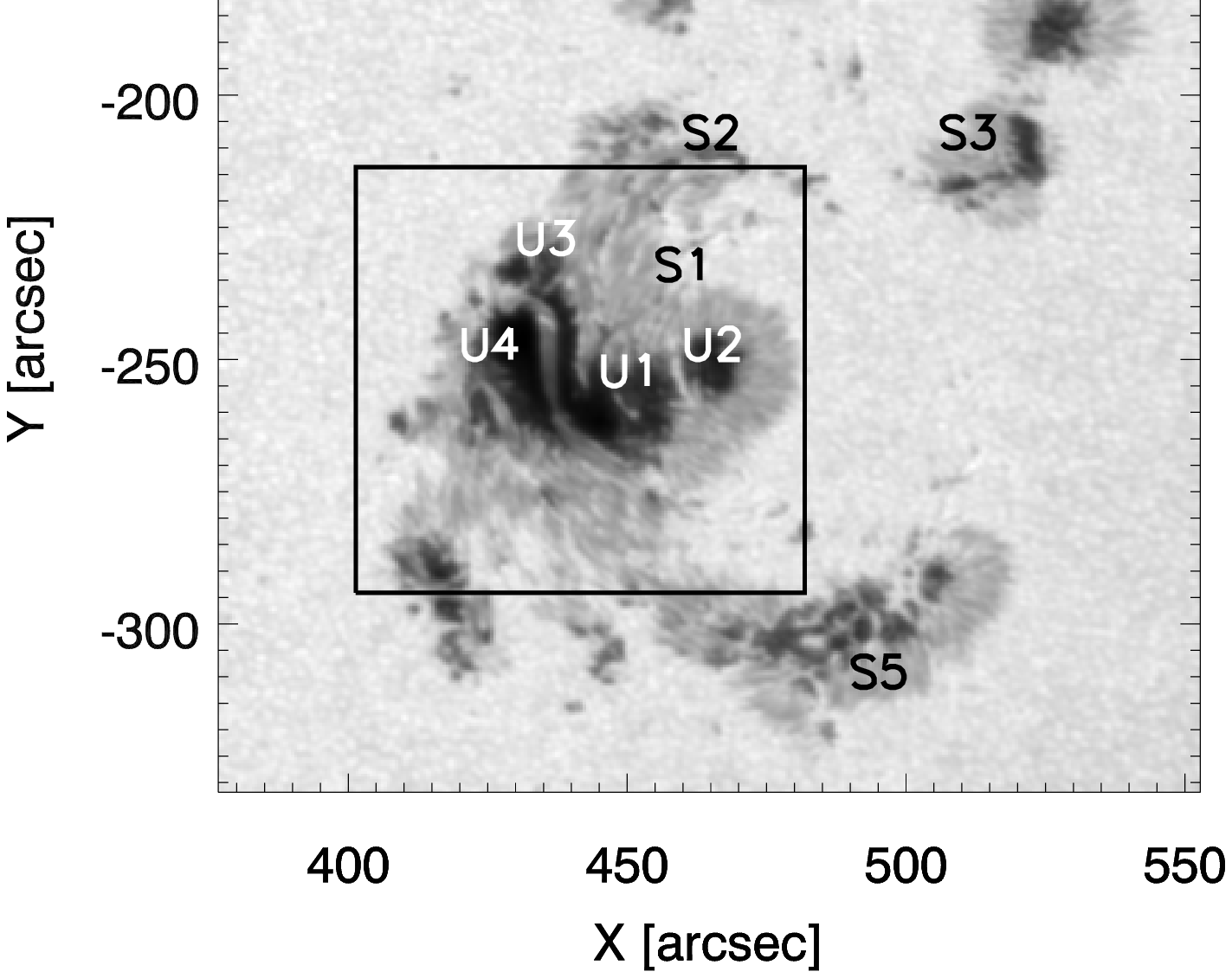}
  \includegraphics[trim=75 100 10 330, clip, scale=0.4]{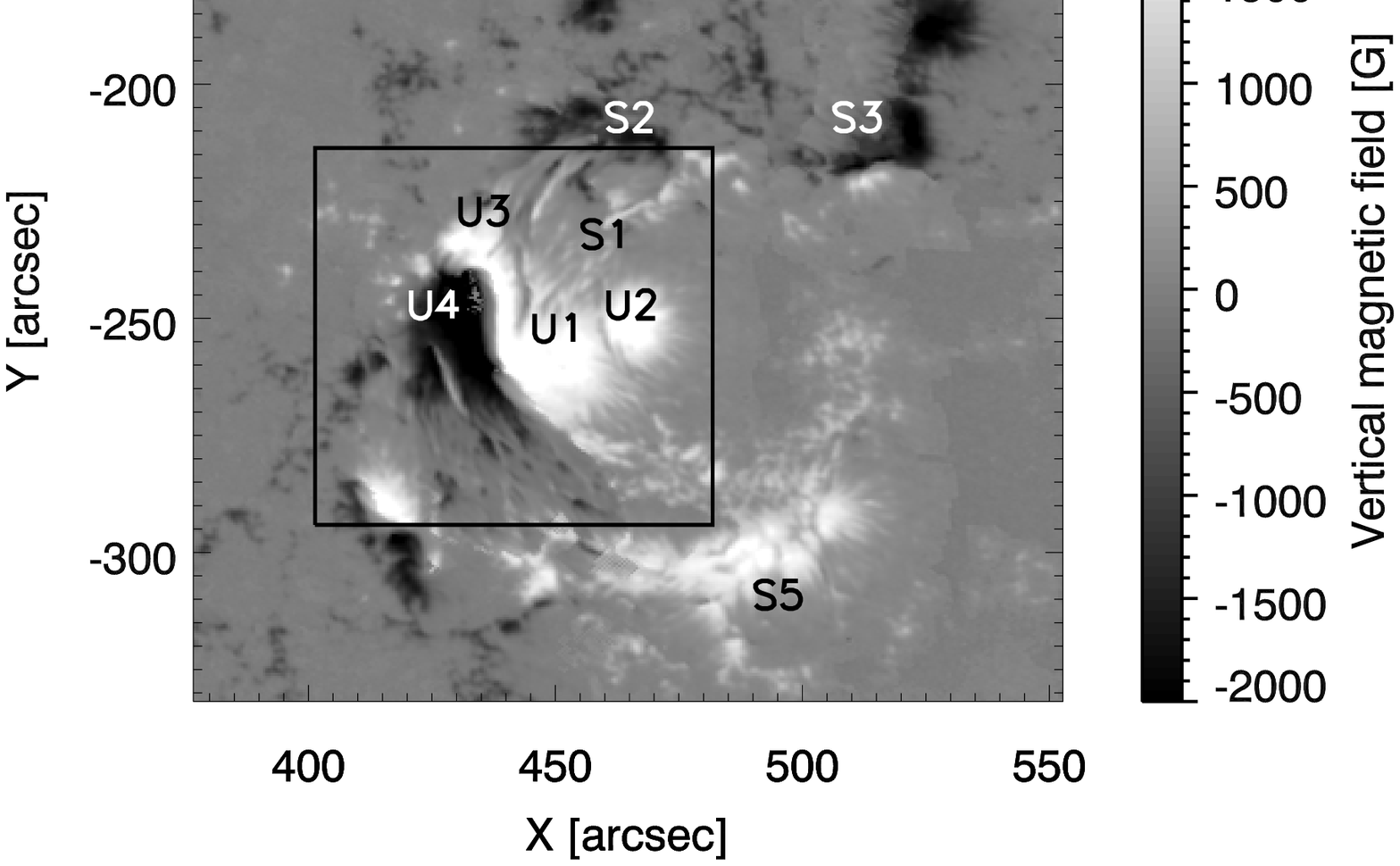}\\
  \caption{AR NOAA 12673 as seen in the continuum filtergram (left panel) and in the vertical component of the vector magnetogram (right panel) obtained in the Fe \small{I} line at 617.3 nm by HMI. The marked boxes highlight the field of view shown in Figures. 3, 5 and 8. \label{Fig2}}
\end{figure*}

\begin{figure*}
  \includegraphics[trim=15 75 410 110, clip, scale=0.39]{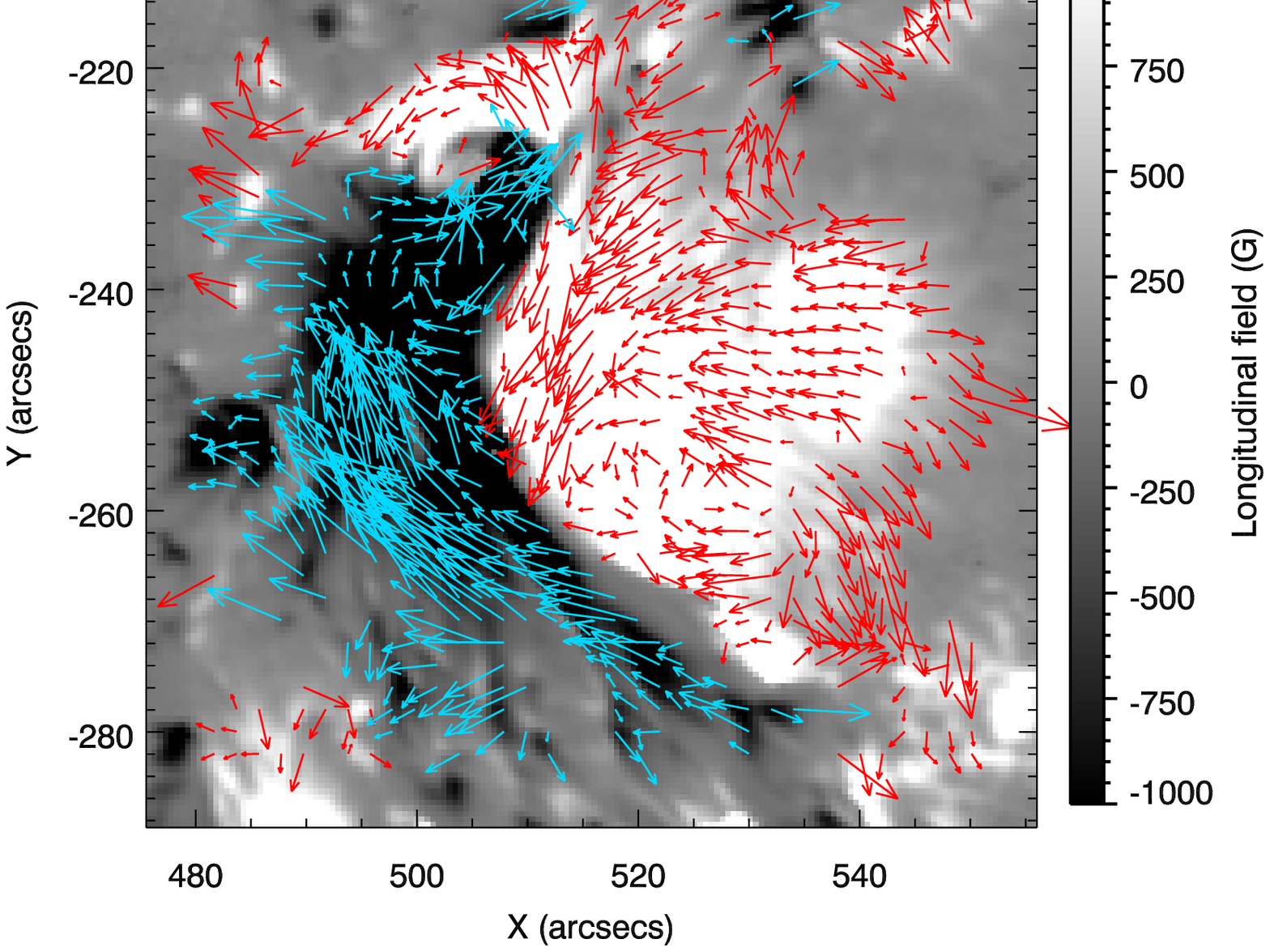}
	 \includegraphics[trim=80 75 320 110, clip, scale=0.39]{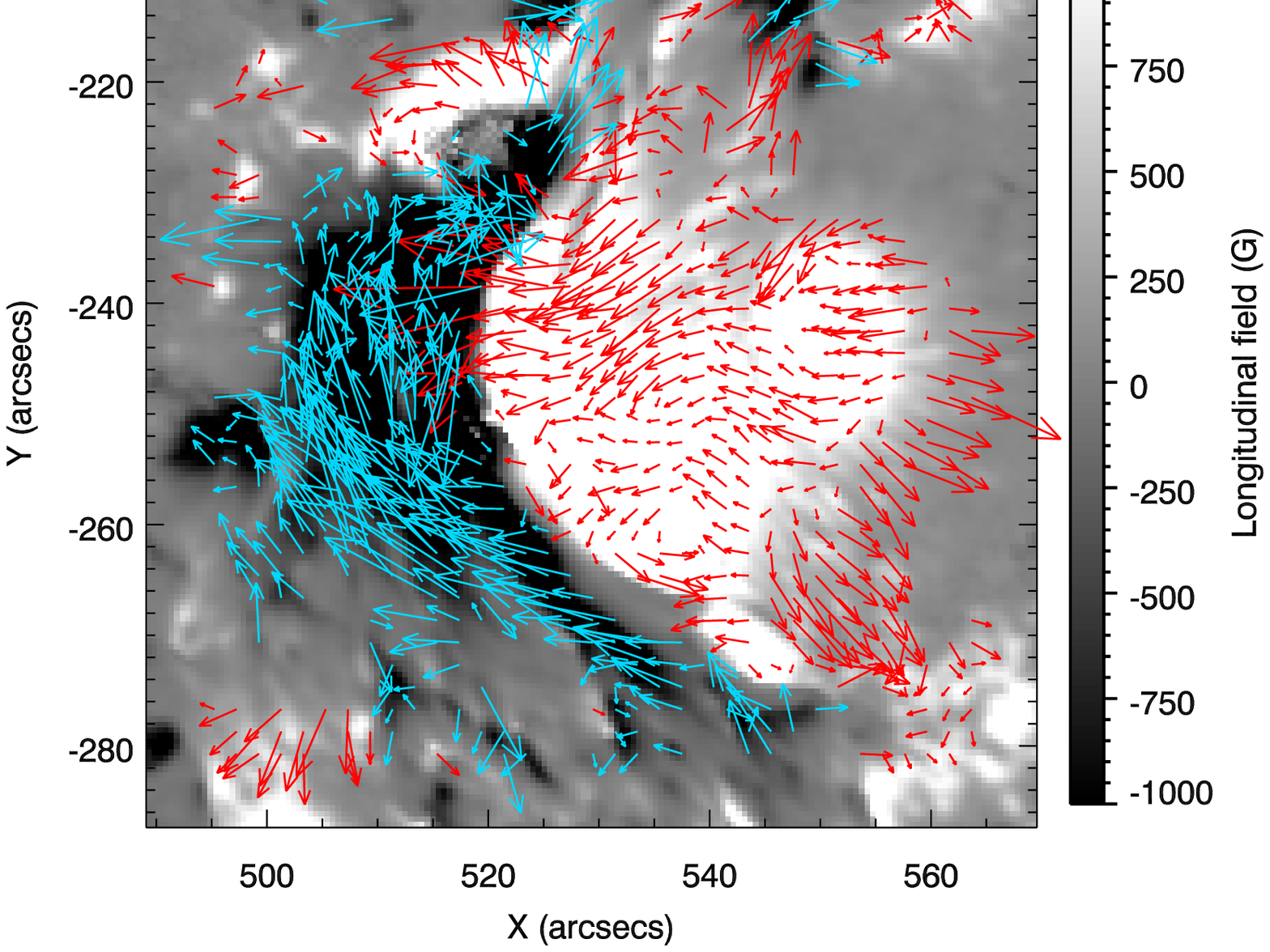}\\
  \caption{Horizontal velocity fields measured over the delta spot of AR NOAA 12673 by means of the DAVE tracking processing method. The background images correspond to the average of the vertical component of the magnetic field measured by HMI during the time interval considered for the velocity computations. The red arrow in the top left corner indicates a scale velocity of 0.5 km s$^{-1}$. \label{Fig3}}
\end{figure*}

\begin{figure*}
  \hspace{-1.3cm}
  \begin{minipage}{2\textwidth}
    \includegraphics[trim=10 50 0 0, clip, height=11cm,width=18.5cm]{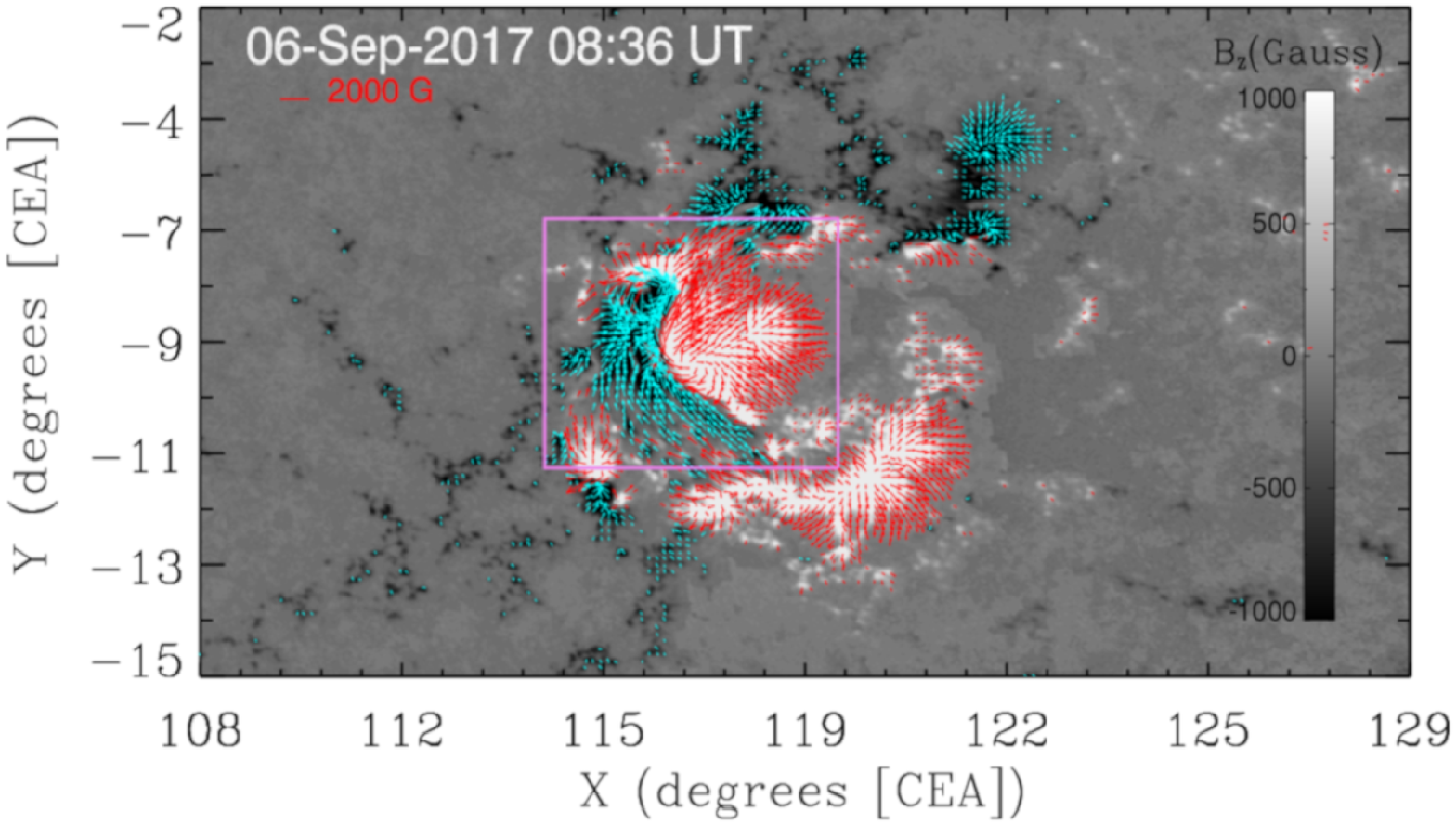}\\
    \includegraphics[trim=-10 0 6 0, clip, scale=0.4]{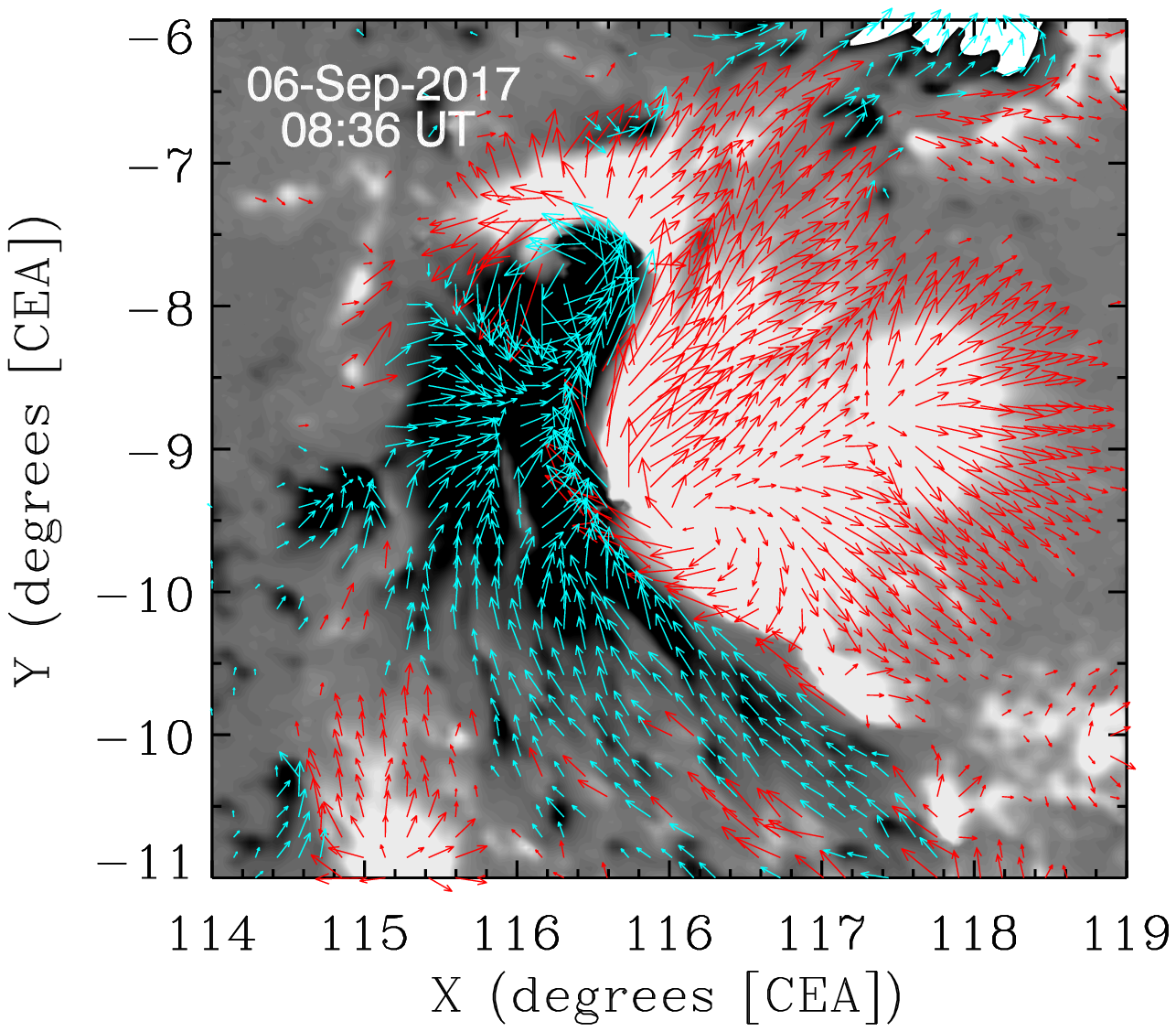}
    \includegraphics[trim=30 0 6 0, clip, scale=0.4]{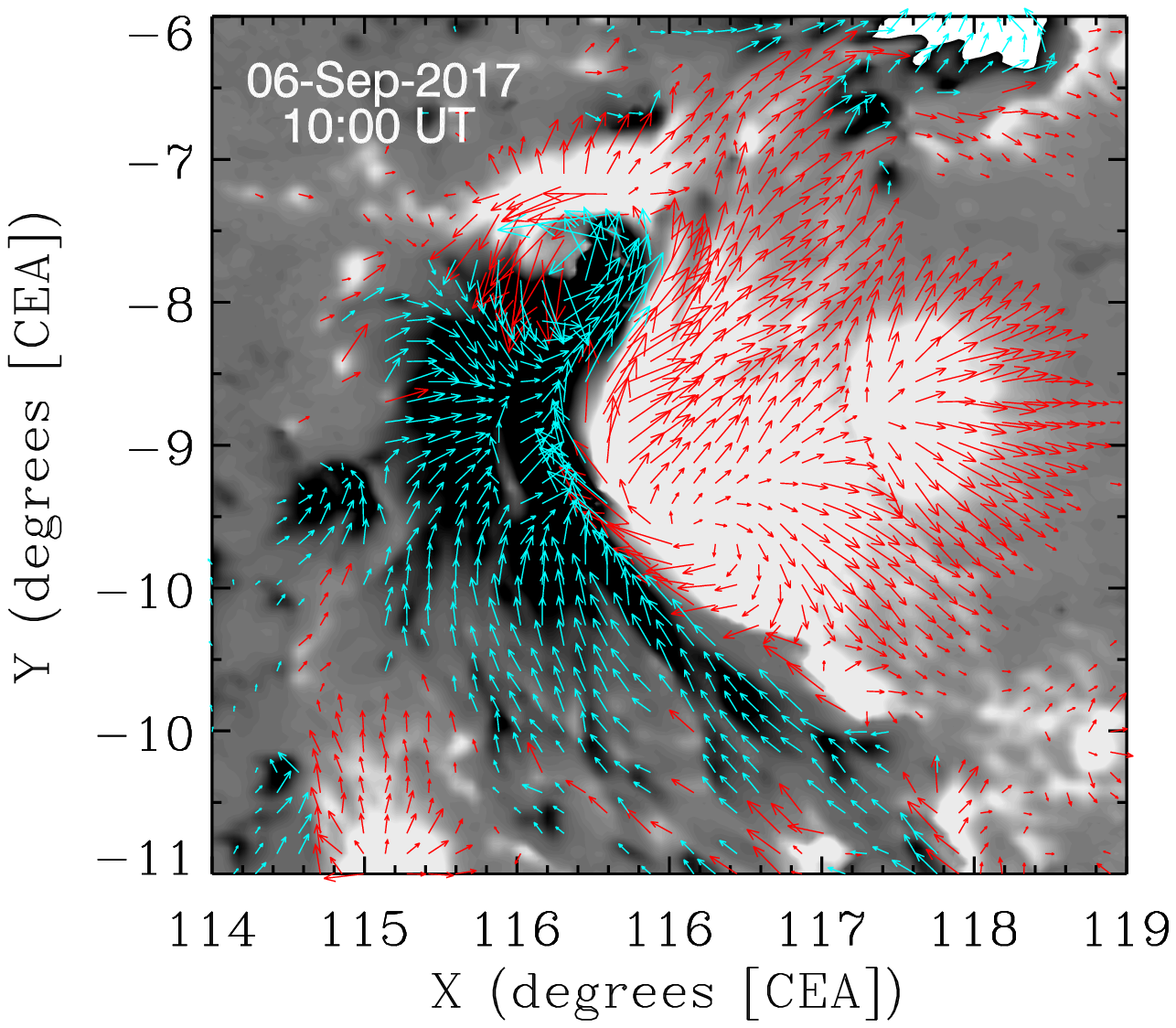}
    \includegraphics[trim=30 0 6 0, clip, scale=0.4]{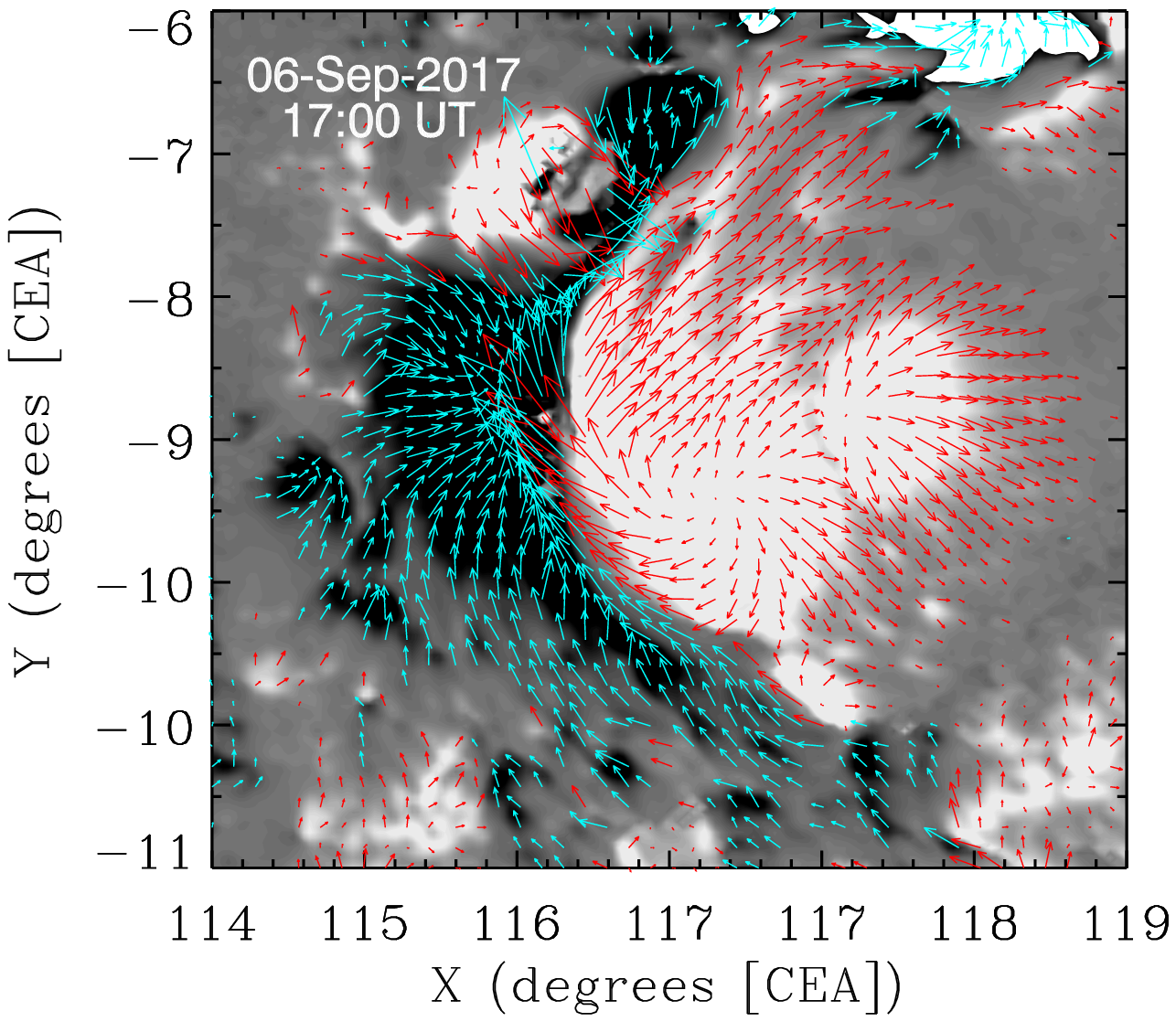}
  \caption{Upper panel: the 08:36 UT transverse field vectors map, overplotted on a B$_z$ background (in $CEA$ coordinates). A cutoff value of $\pm$200 G for the transverse fields is adopted in the computations. Bottom panels: zoomed views of the region of interest ({\it i.e.} enclosed by a box in top panel),  before the first peak (at 08:36 UT), between the two peaks (at 10:00 UT) and hours after the second peak (at 17:00 UT). The red arrows indicate the transverse magnetic fields of positive polarity, while the cyan arrows denote the transverse magnetic fields of negative polarity. Note the strong shear topology along the PIL, as well as the evident counterclockwise rotation of the north-eastern component of the negative polarity.\label{Fig4}}
 \end{minipage}
\end{figure*}

\begin{figure*}
  \includegraphics[trim=5 100 170 300, clip, scale=0.4]{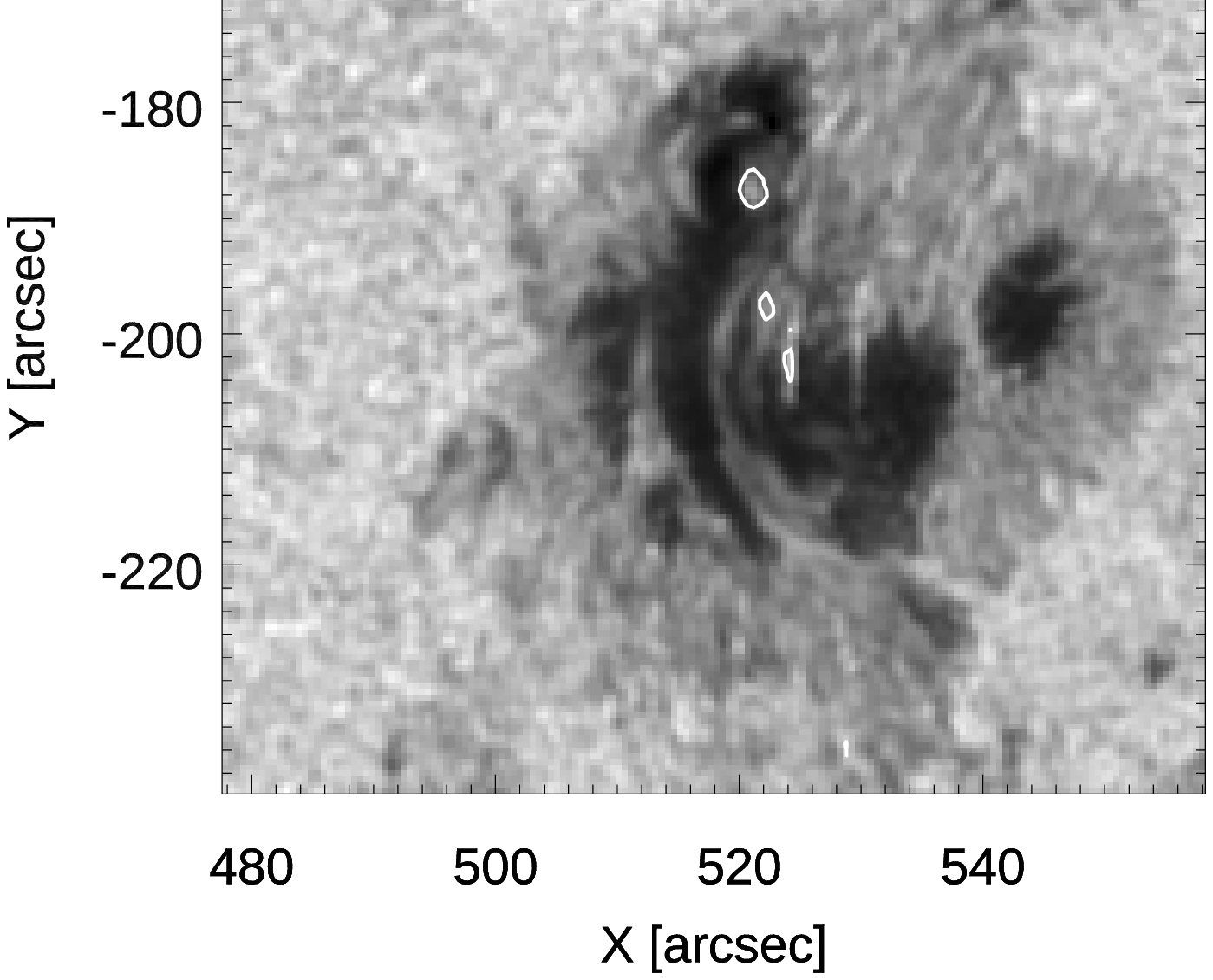}
  \includegraphics[trim=75 100 10 330, clip, scale=0.4]{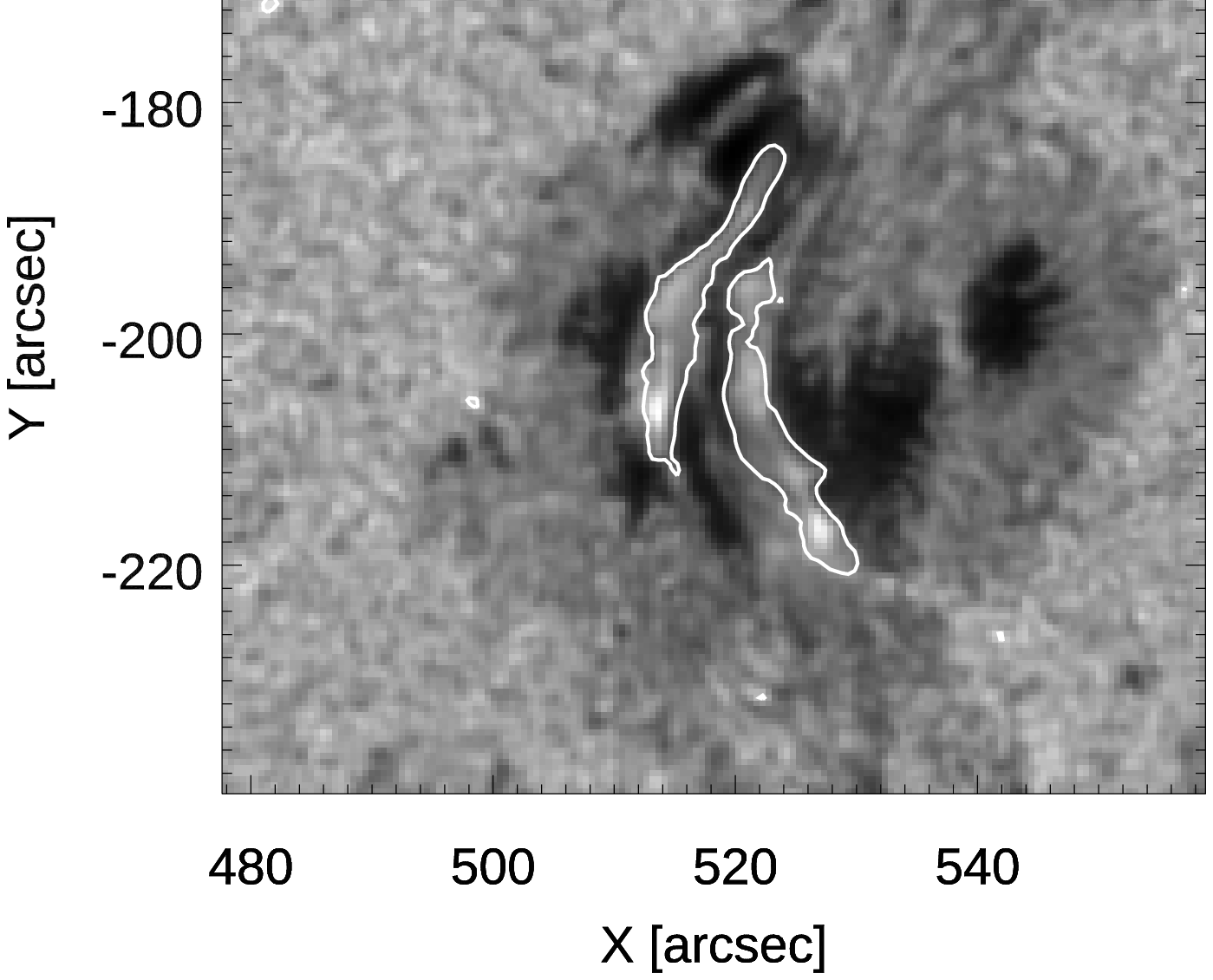}\\
  \caption{Continuum filtergrams, obtained in the Fe \small{I} line at 617.3 nm, by HMI during the peaks of the two X-GOES class flares occurring in AR NOAA 12673. The contours highlight the locations of the white light emissions during the flares.\label{Fig5}}
\end{figure*}

\begin{figure*}
  \includegraphics[trim=5 105 170 310, clip, scale=0.4]{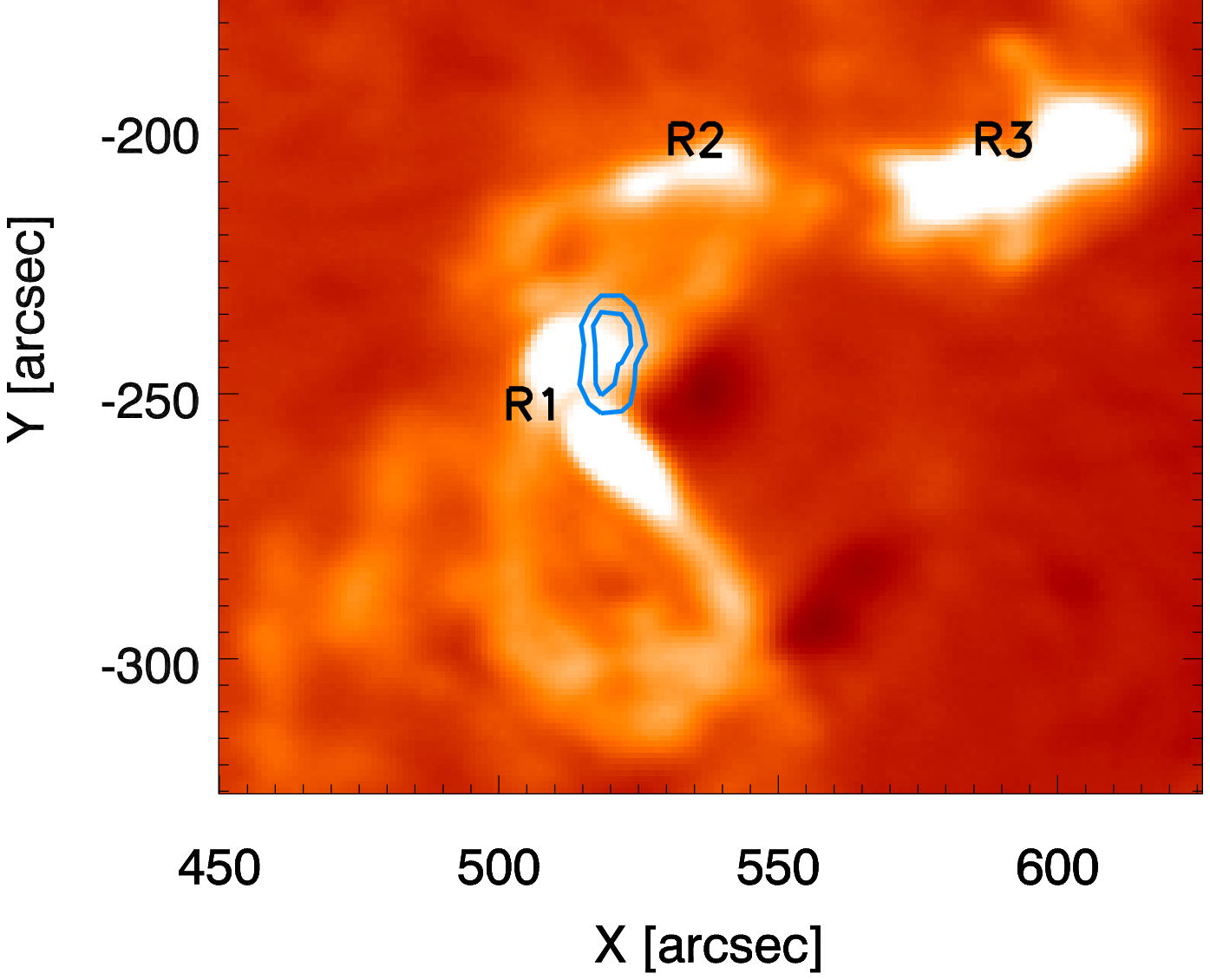}
  \includegraphics[trim=75 105 10 310, clip, scale=0.4]{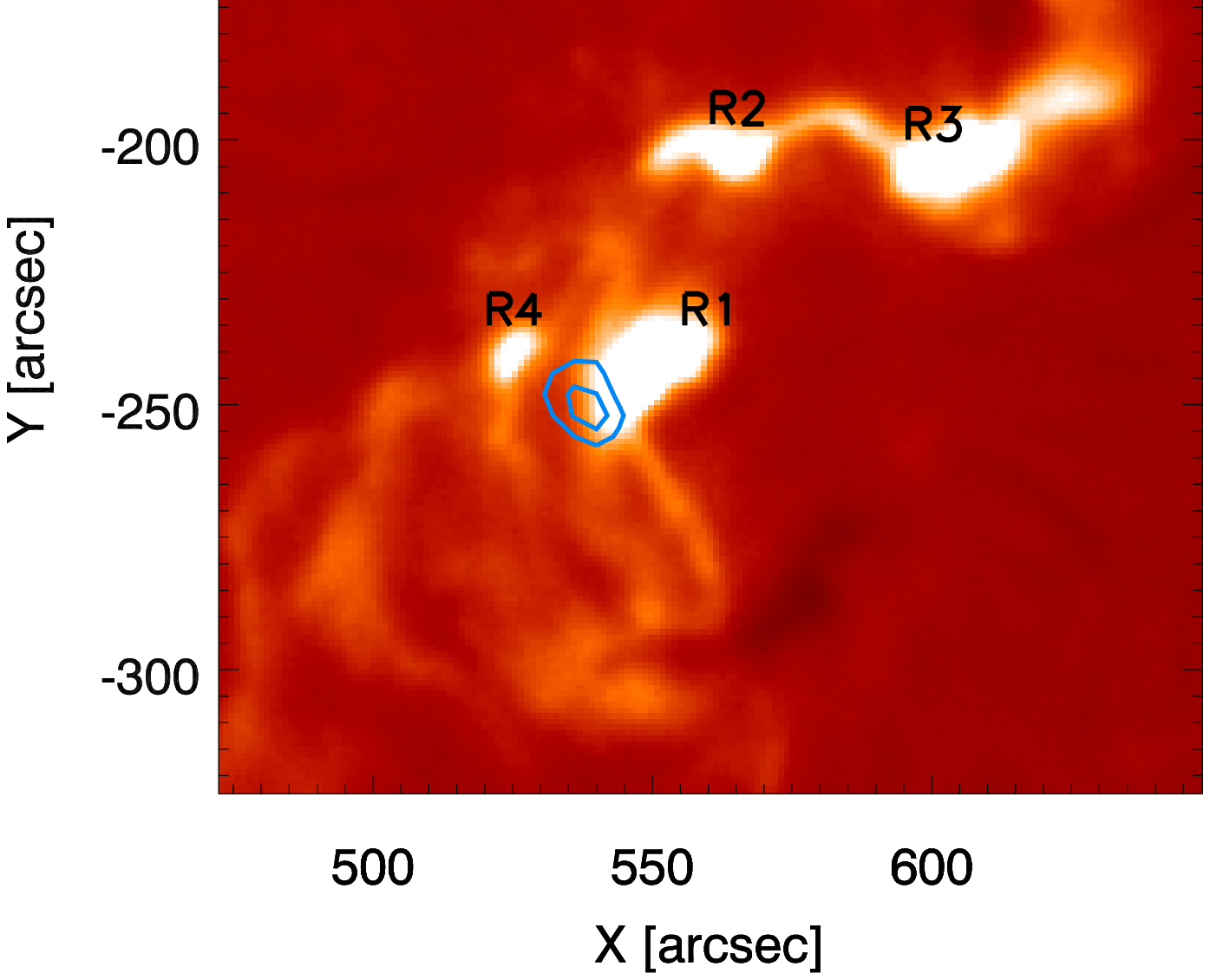}\\
	\includegraphics[trim=5 105 170 310, clip, scale=0.4]{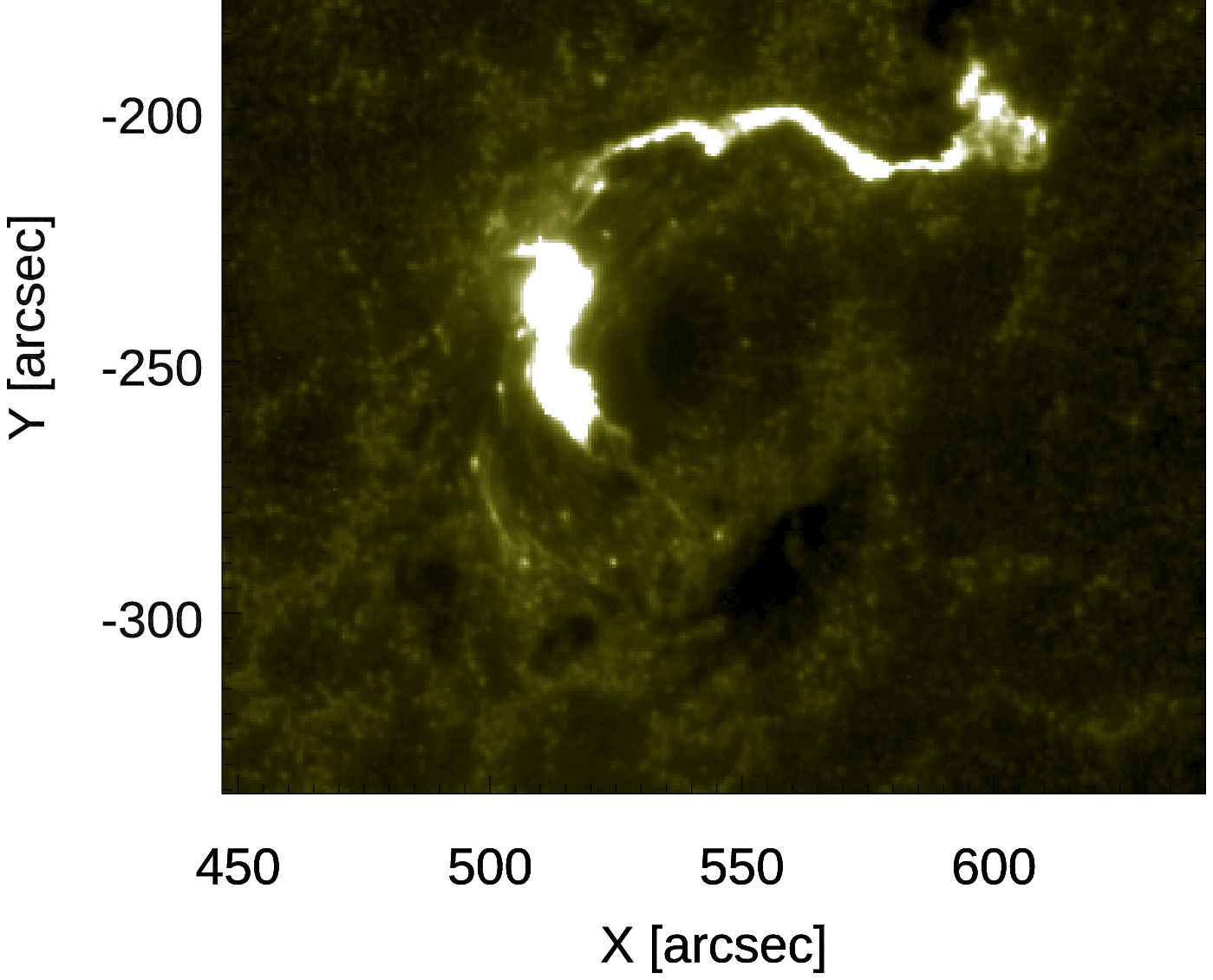}
  \includegraphics[trim=75 105 10 310, clip, scale=0.4]{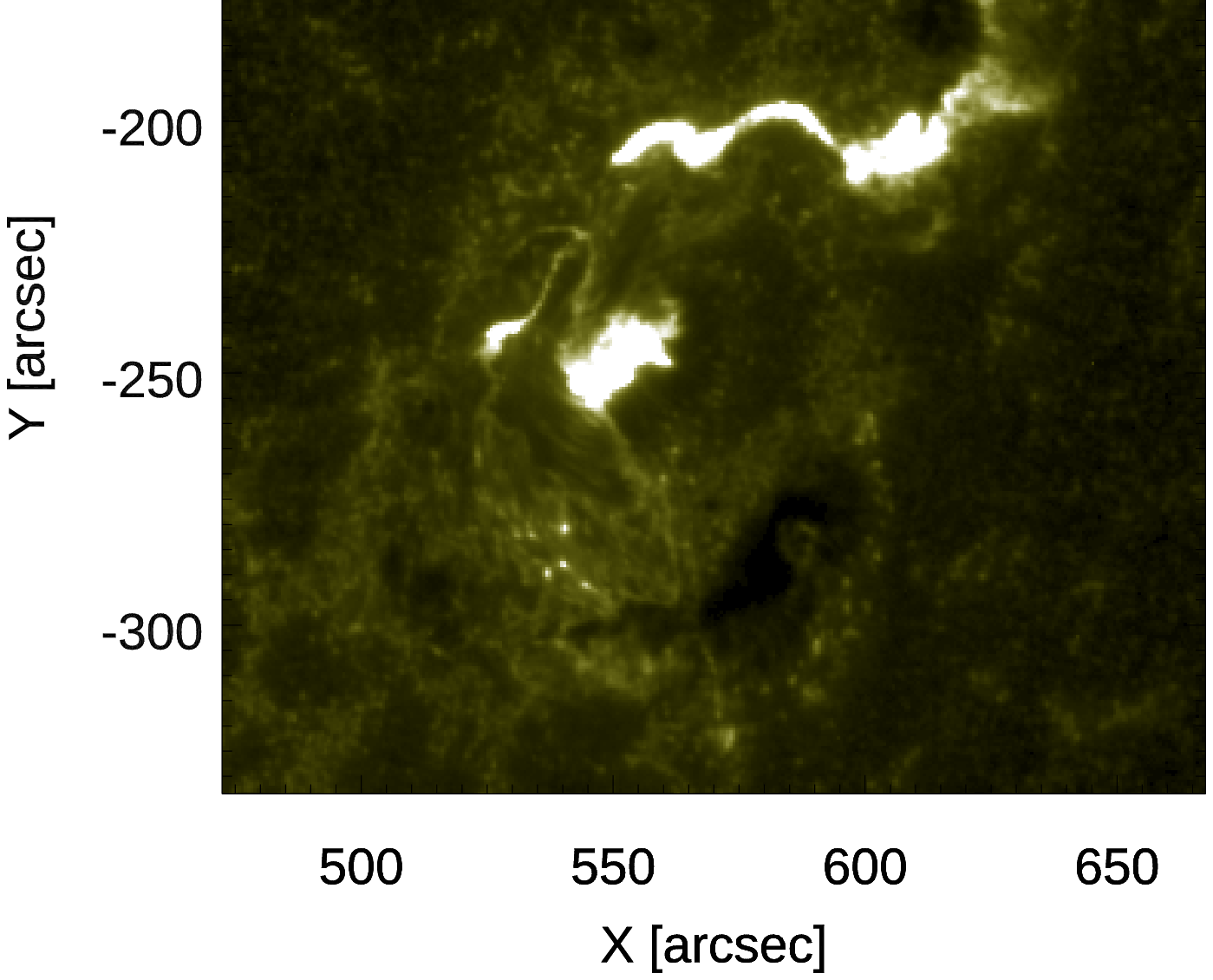}\\
  \caption{Upper panels: AR NOAA 12673 in the H$\alpha$ line during the main phase of the X2.2 (left panel) and the X9.3 (right panel) GOES class flares. The reported overlaid contours indicate 70\% and 90\% of the flux peak at 12-25 keV taken by RHESSI around the peaks of the WLFs. The shown labels indicate the different kernels of the ribbons observed at chromospheric level.
  Lower panels: AR NOAA 12673 as observed at 1600 \AA{} by AIA during the main phases of the X2.2 (left panel) and the X9.3 (right panel) events.
  \label{Fig6}}
\end{figure*}

\begin{figure*}
  \includegraphics[trim=5 100 170 300, clip, scale=0.4]{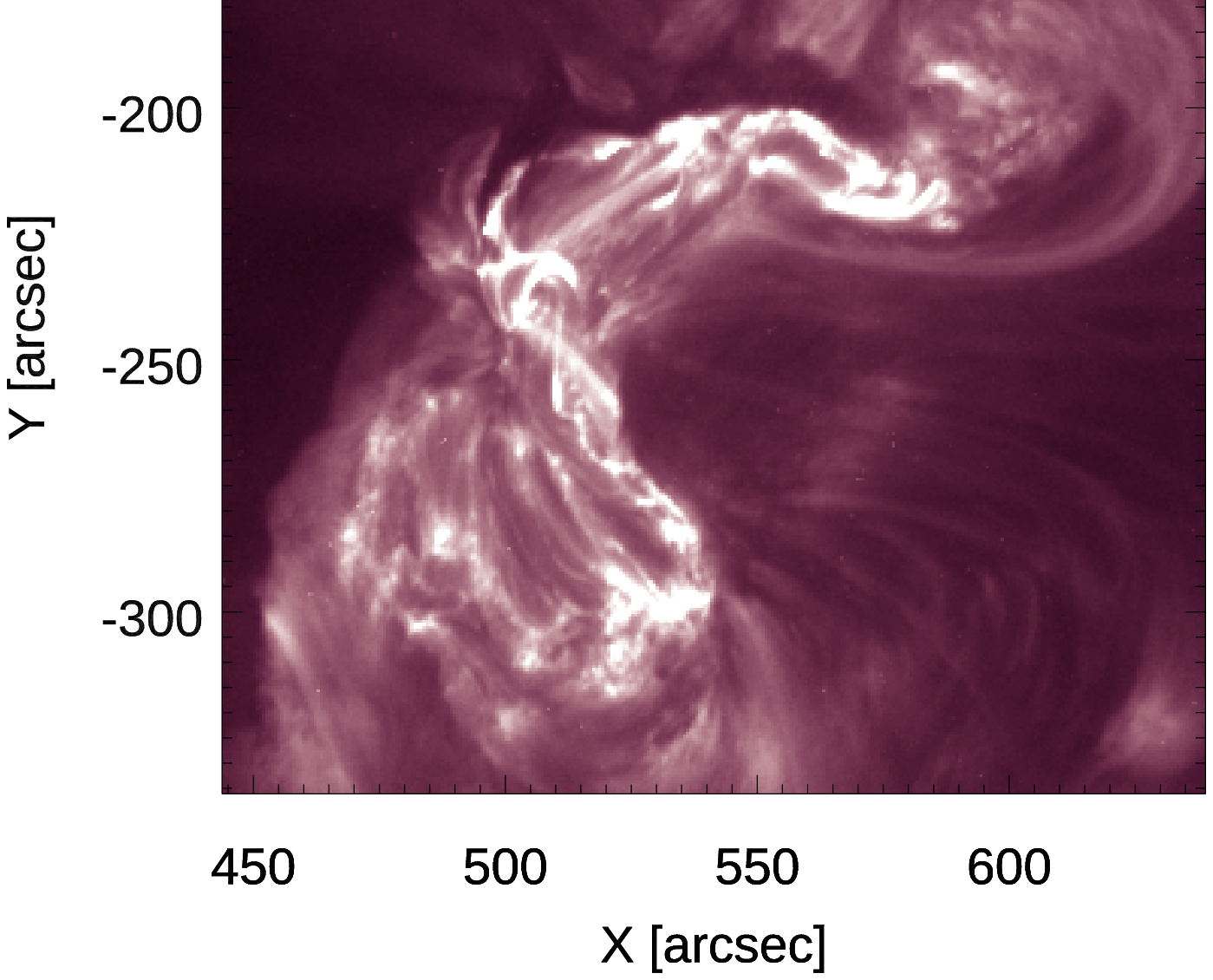}
  \includegraphics[trim=75 100 10 330, clip, scale=0.4]{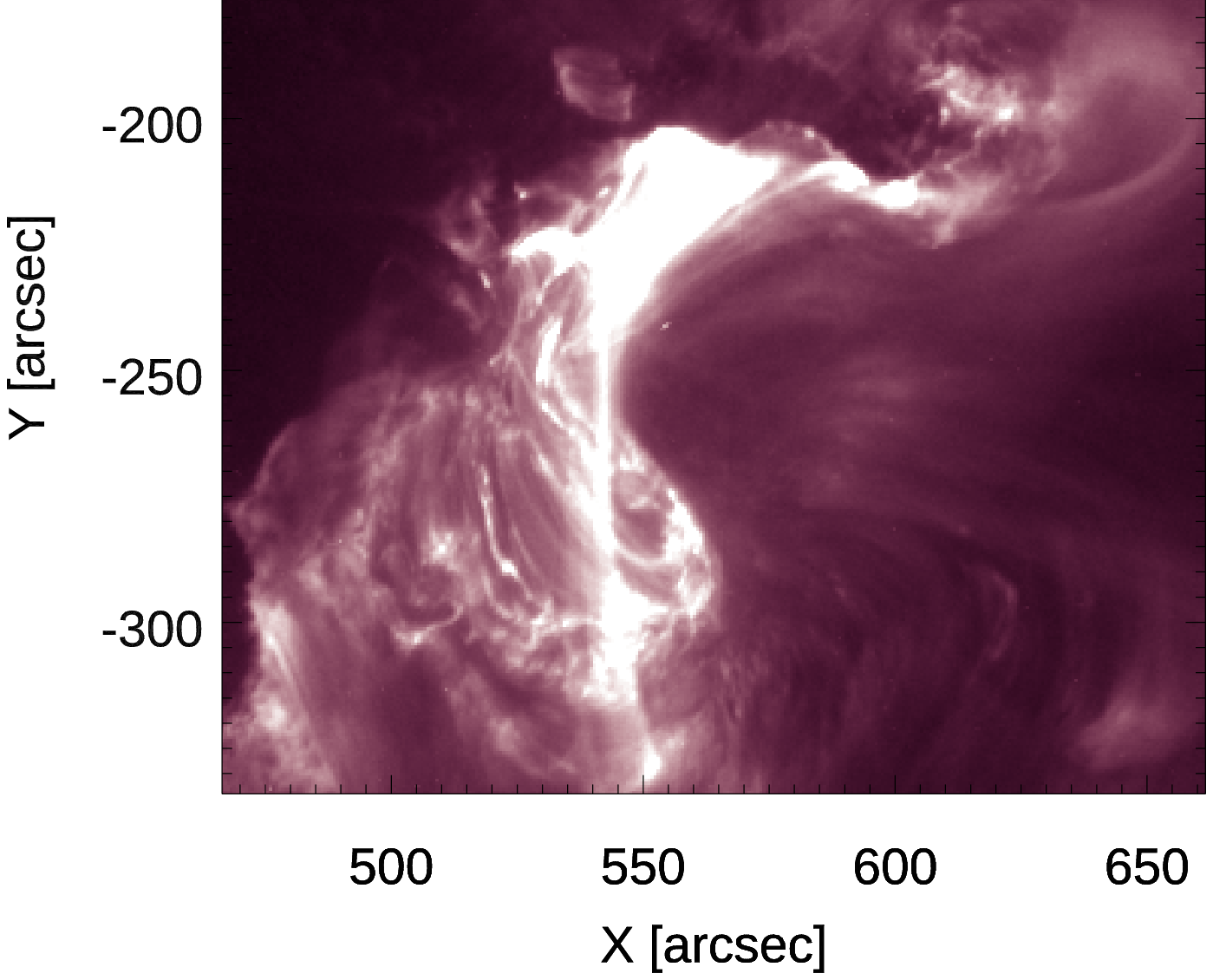}\\
  \caption{\bf AR NOAA 12673 images taken at 211 \AA{} by AIA before the beginning of the WLFs, at 8:39 UT (left panel) and at 11:45 UT (right panel). Note the appearance of an helical structure crossing the AR (refer to the text for more details).\label{Fig7}}
\end{figure*}

\begin{figure*}
  \includegraphics[trim=5 105 170 310, clip, scale=0.4]{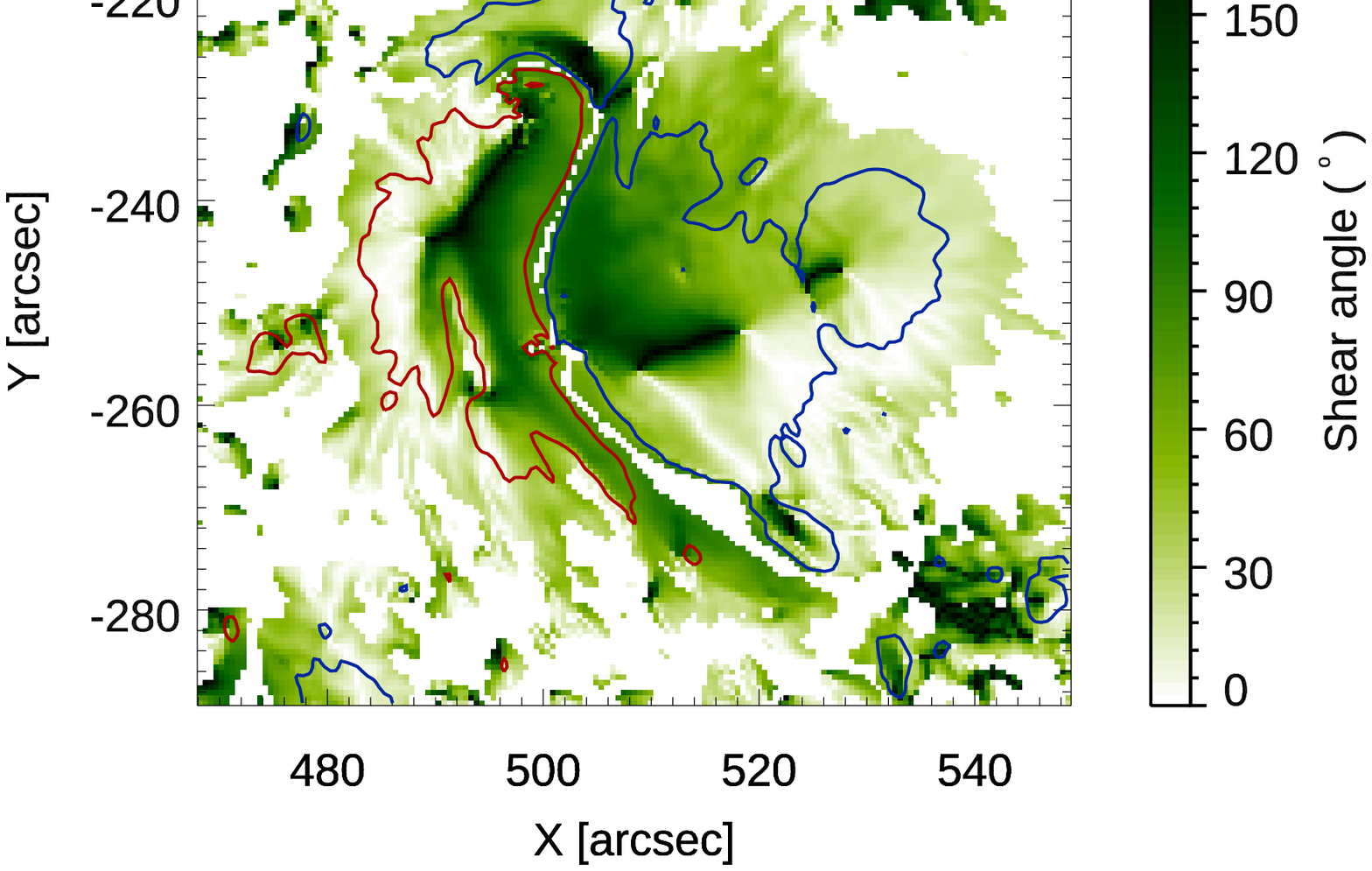}
  \includegraphics[trim=75 105 10 310, clip, scale=0.4]{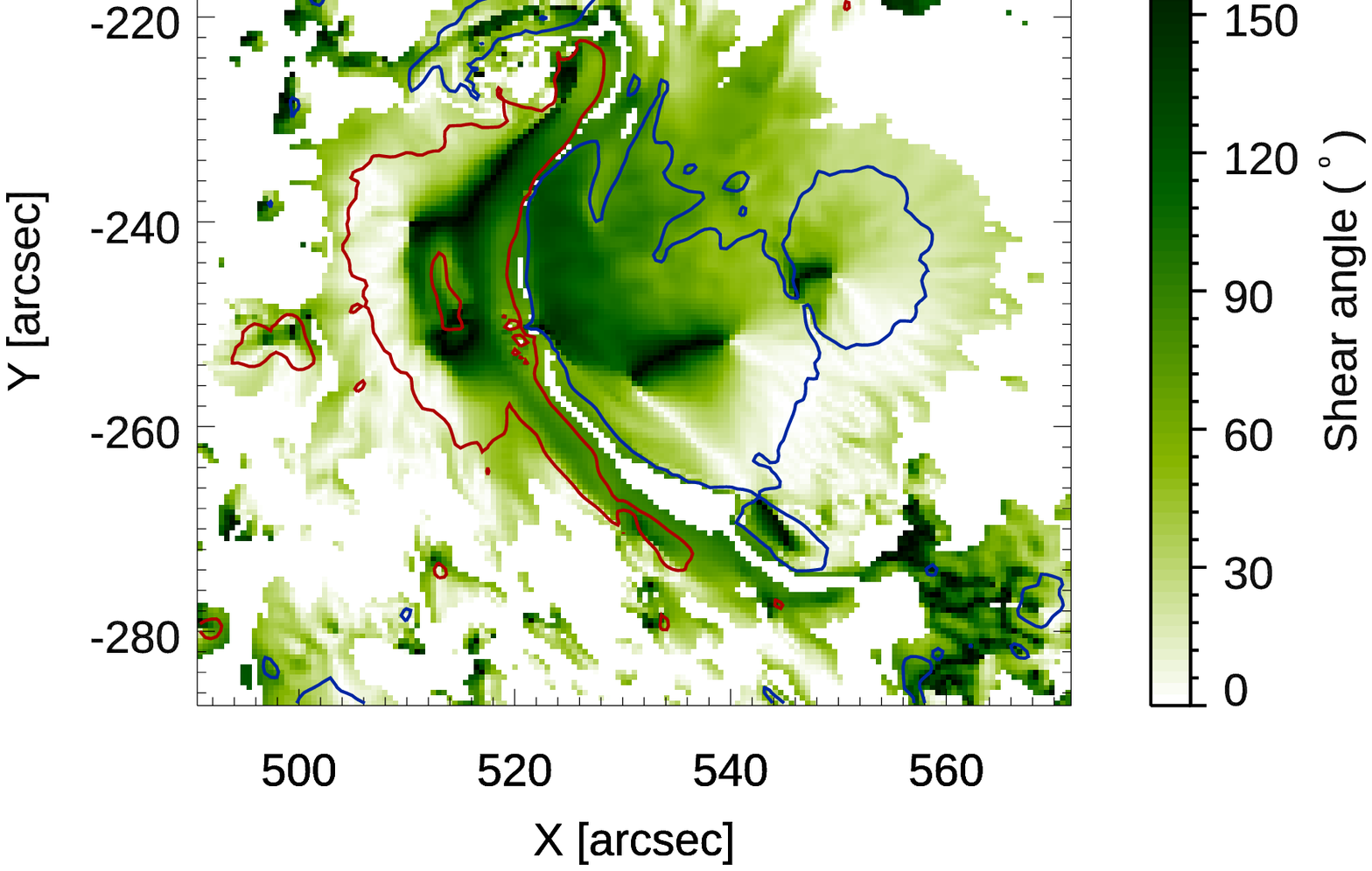}\\
	\includegraphics[trim=5 105 170 310, clip, scale=0.4]{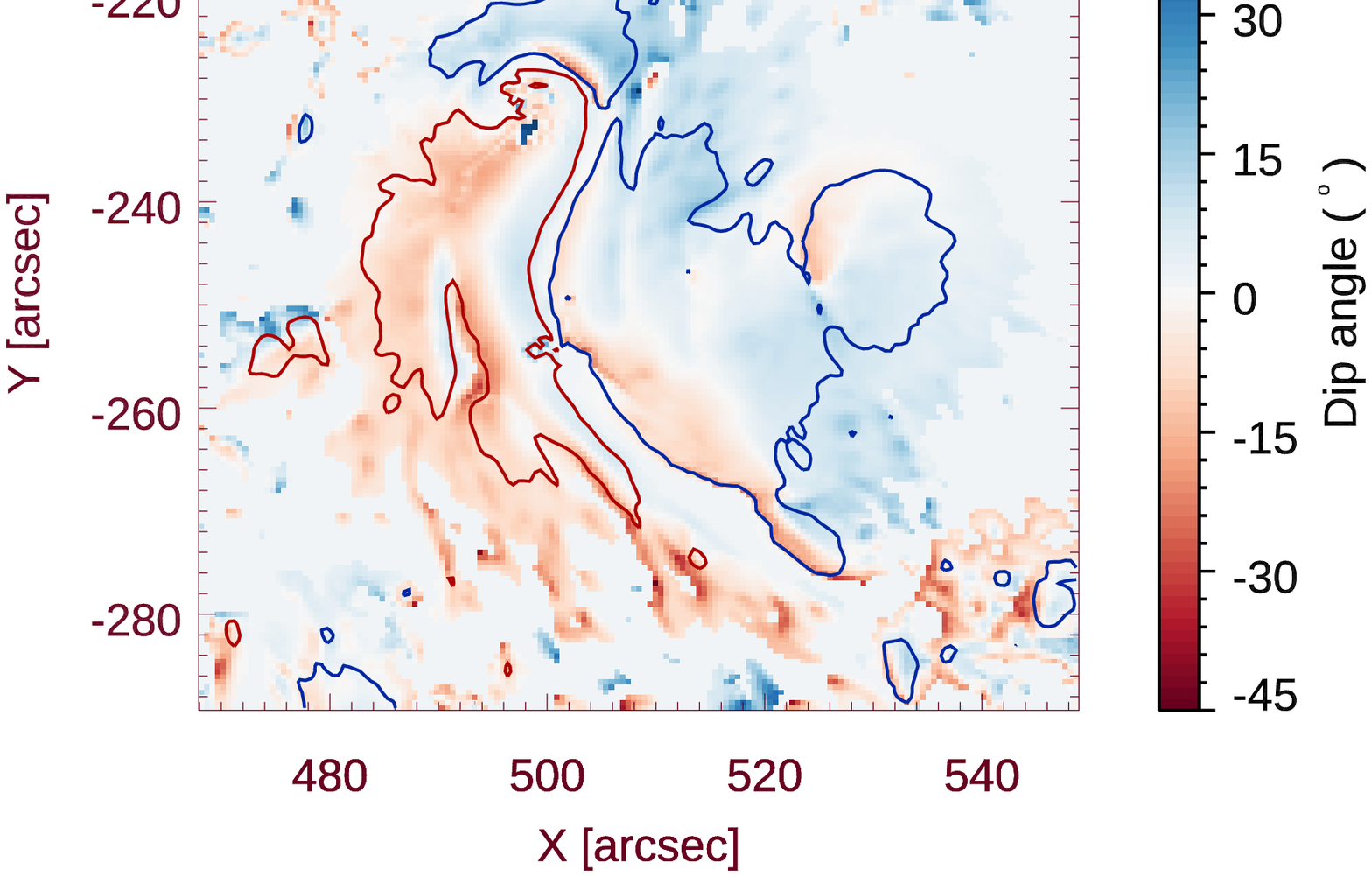}
  \includegraphics[trim=75 105 10 310, clip, scale=0.4]{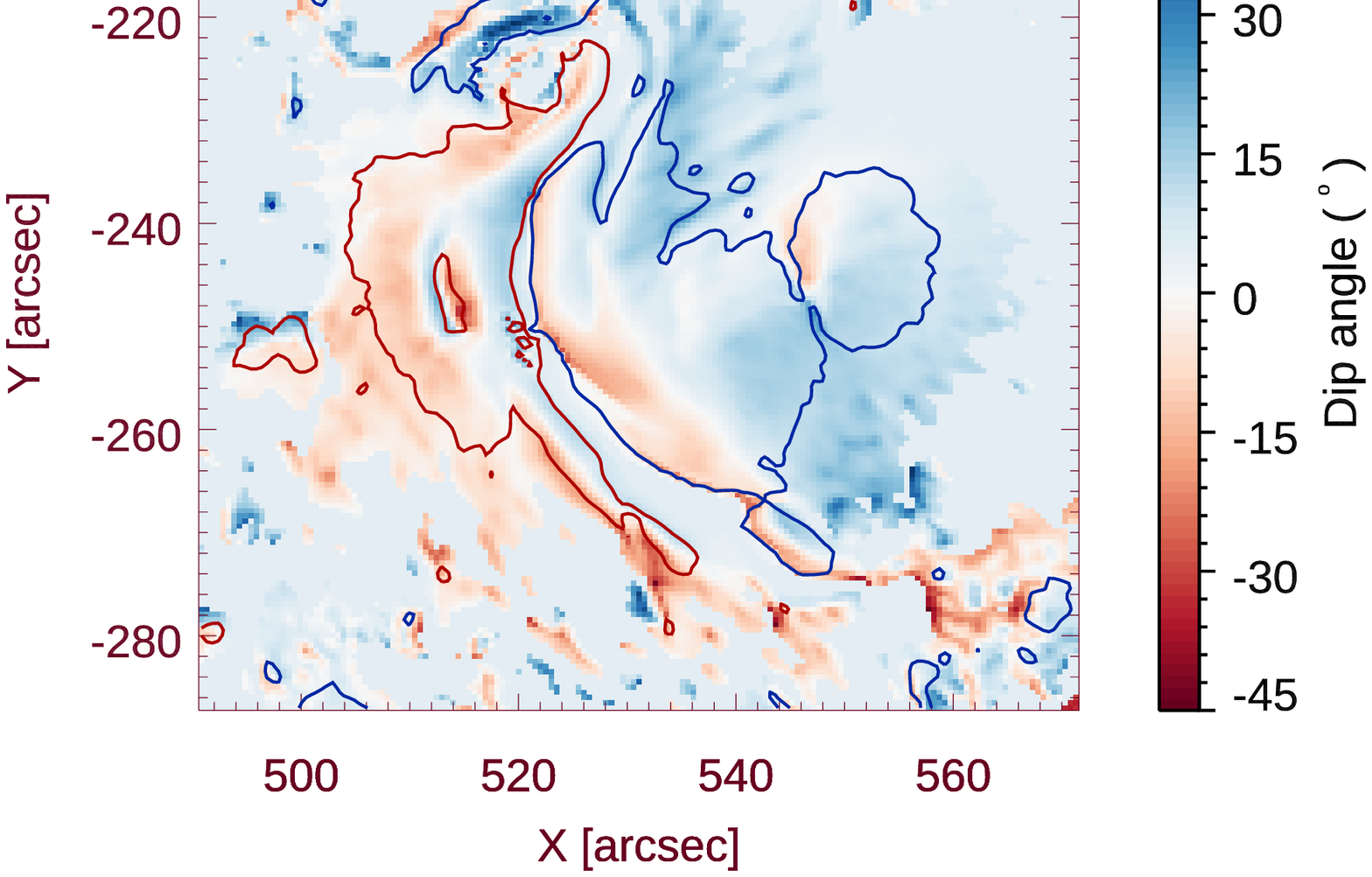}\\
  \caption{\bf Maps of the shear (upper panels) and dip (lower panels) angles of AR NOAA 12673, both at 8:46 UT (left panels) and at 11:46 UT (right panels). The blue (red) contours indicate a longitudinal field of +1000G (-1000 G), respectively.\label{Fig8}}
\end{figure*}

\begin{figure*}
    \includegraphics[trim=0 0 0 0, clip, scale=0.27]{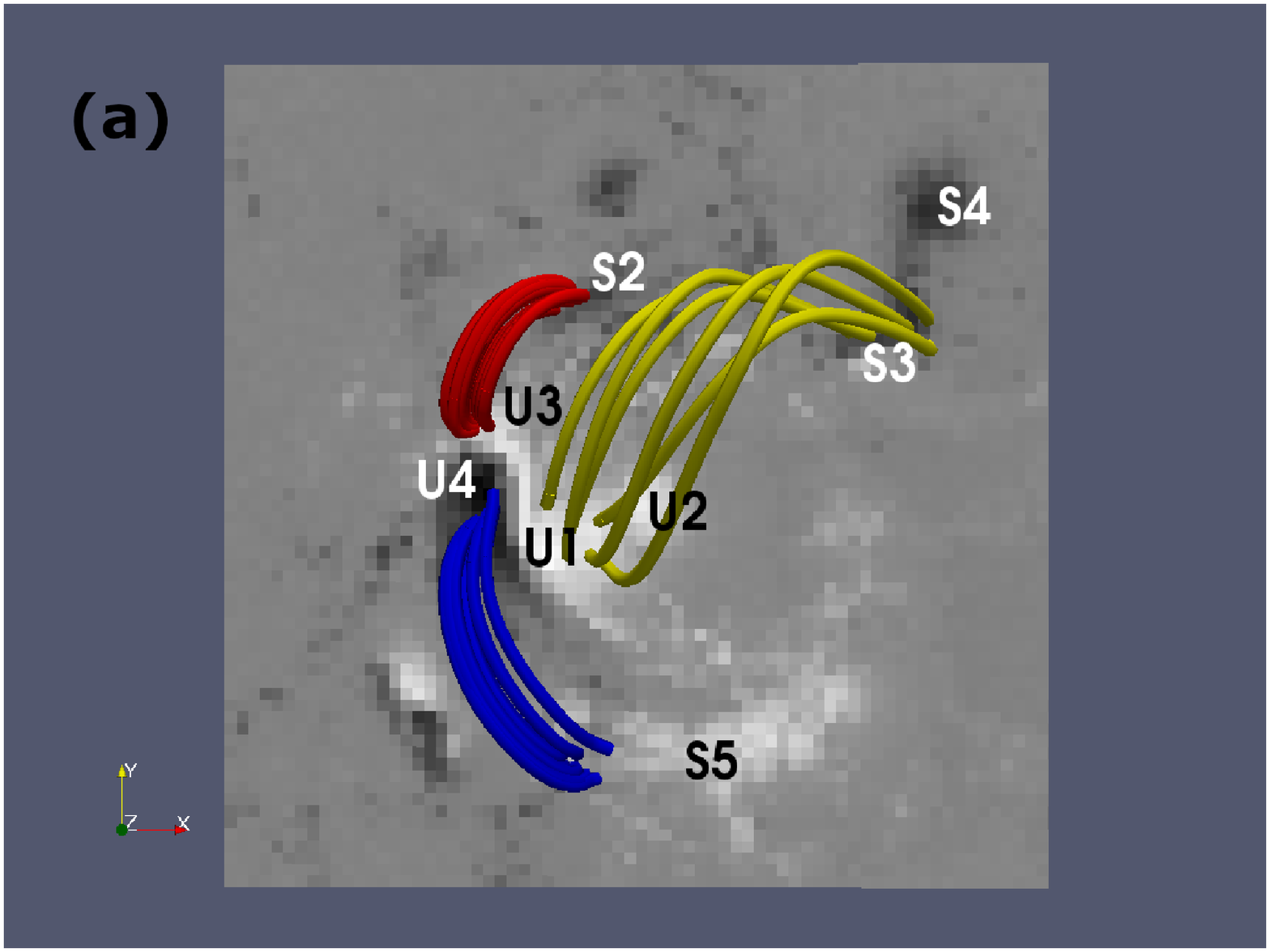}
    \includegraphics[trim=0 0 0 0, clip, scale=0.27]{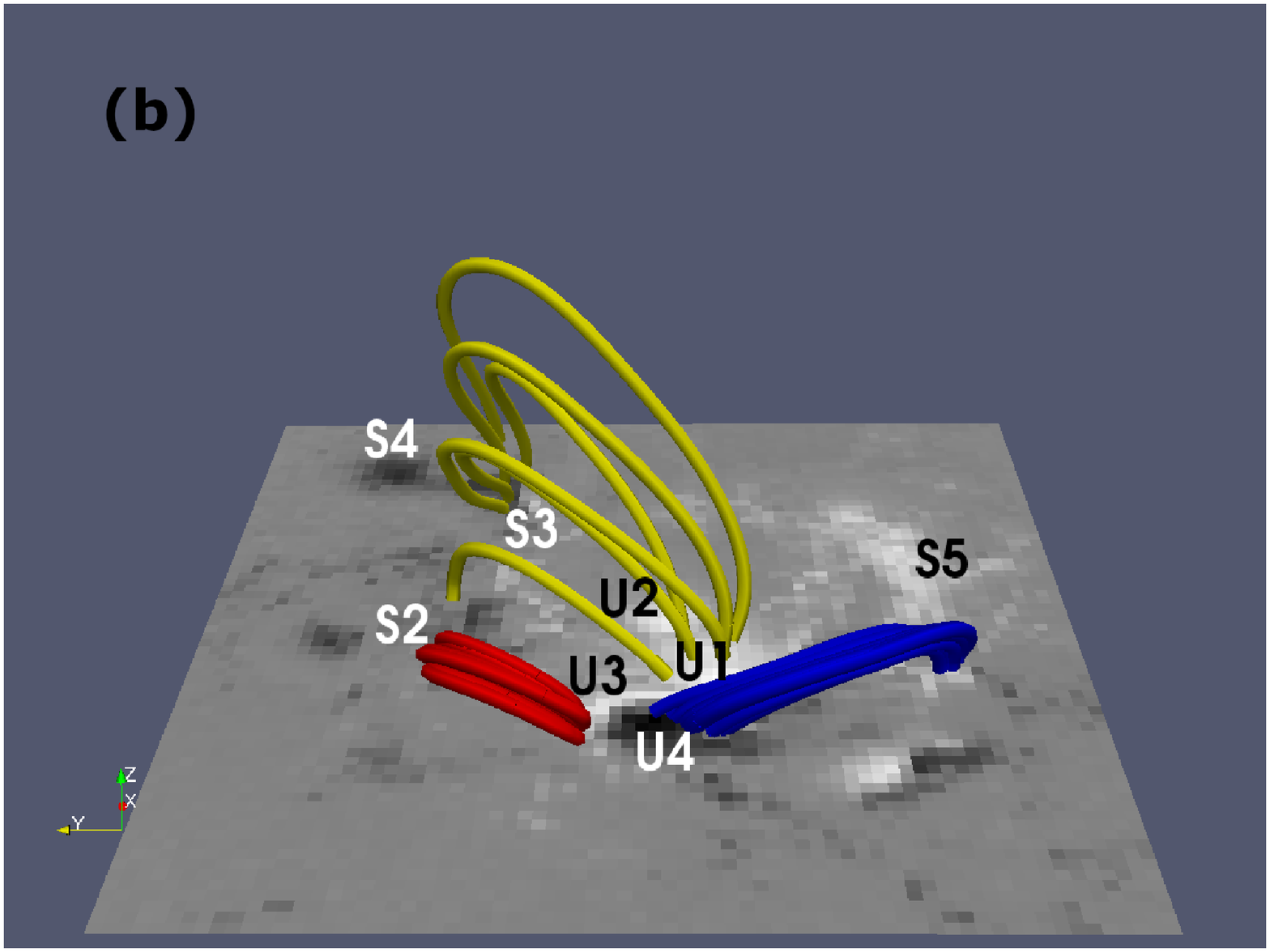}\\
	\includegraphics[trim=0 0 0 0, clip, scale=0.27]{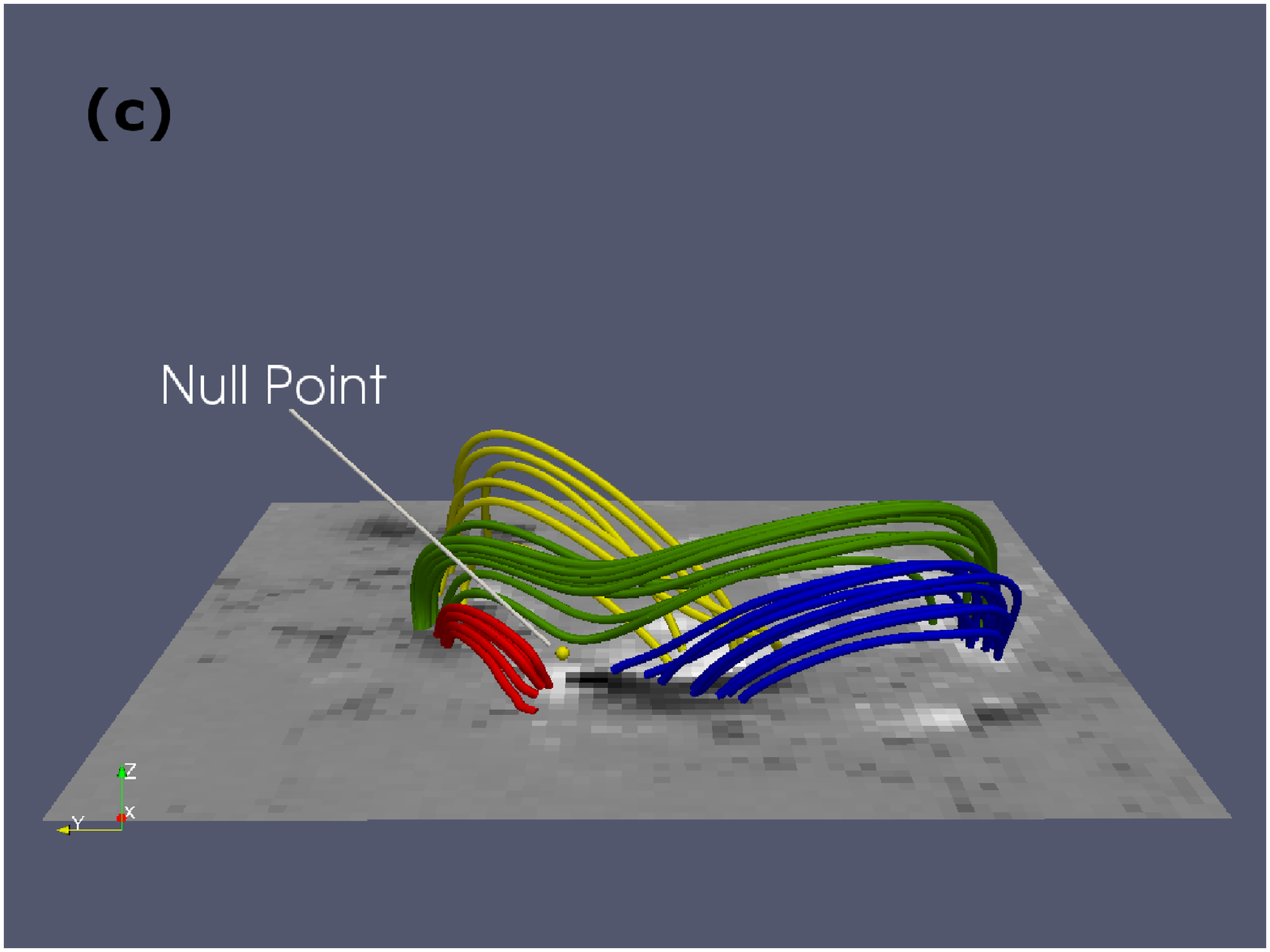}
    \includegraphics[trim=0 0 0 0, clip, scale=0.27]{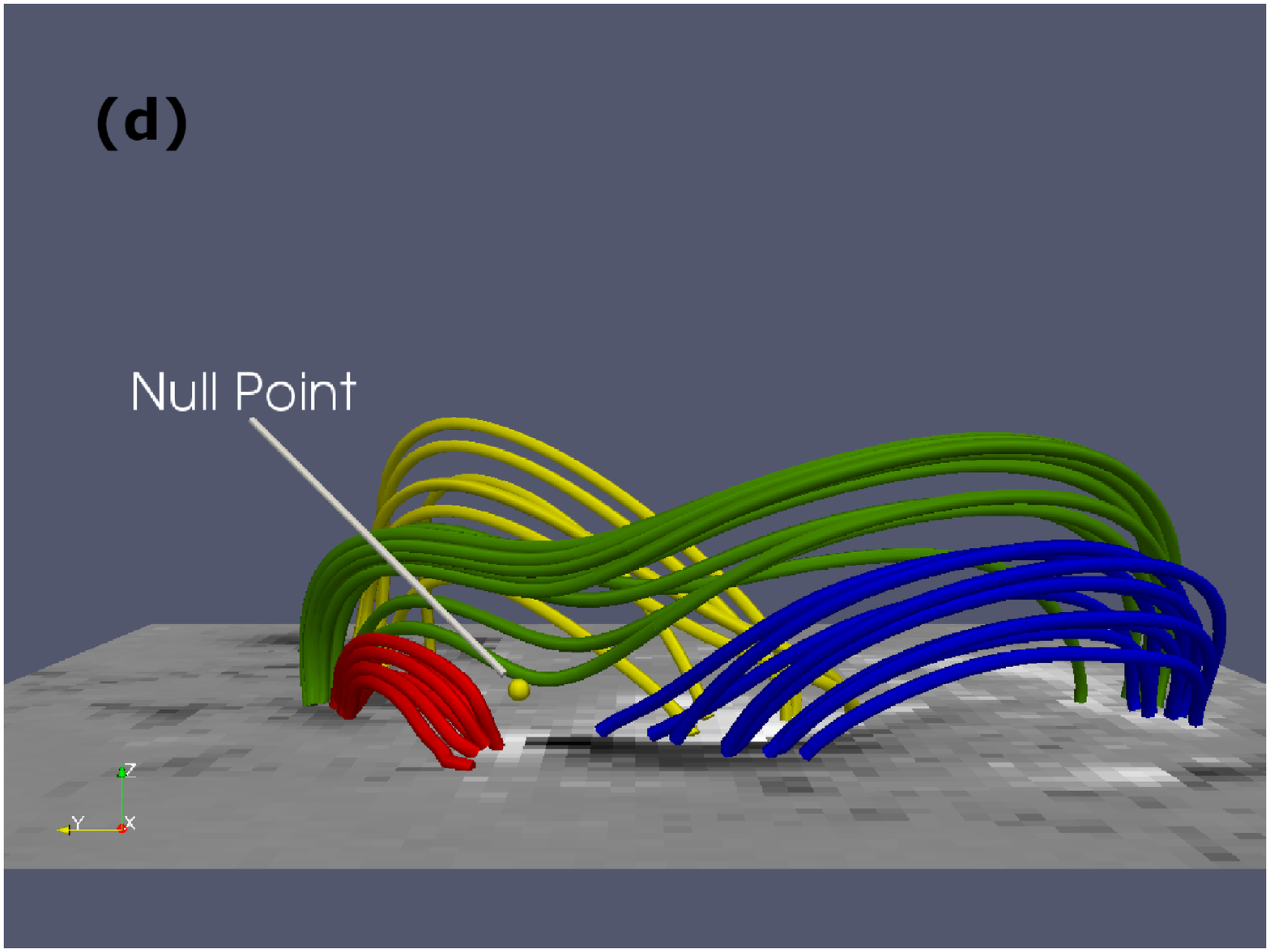}\\
	\includegraphics[trim=0 0 0 0, clip, scale=0.27]{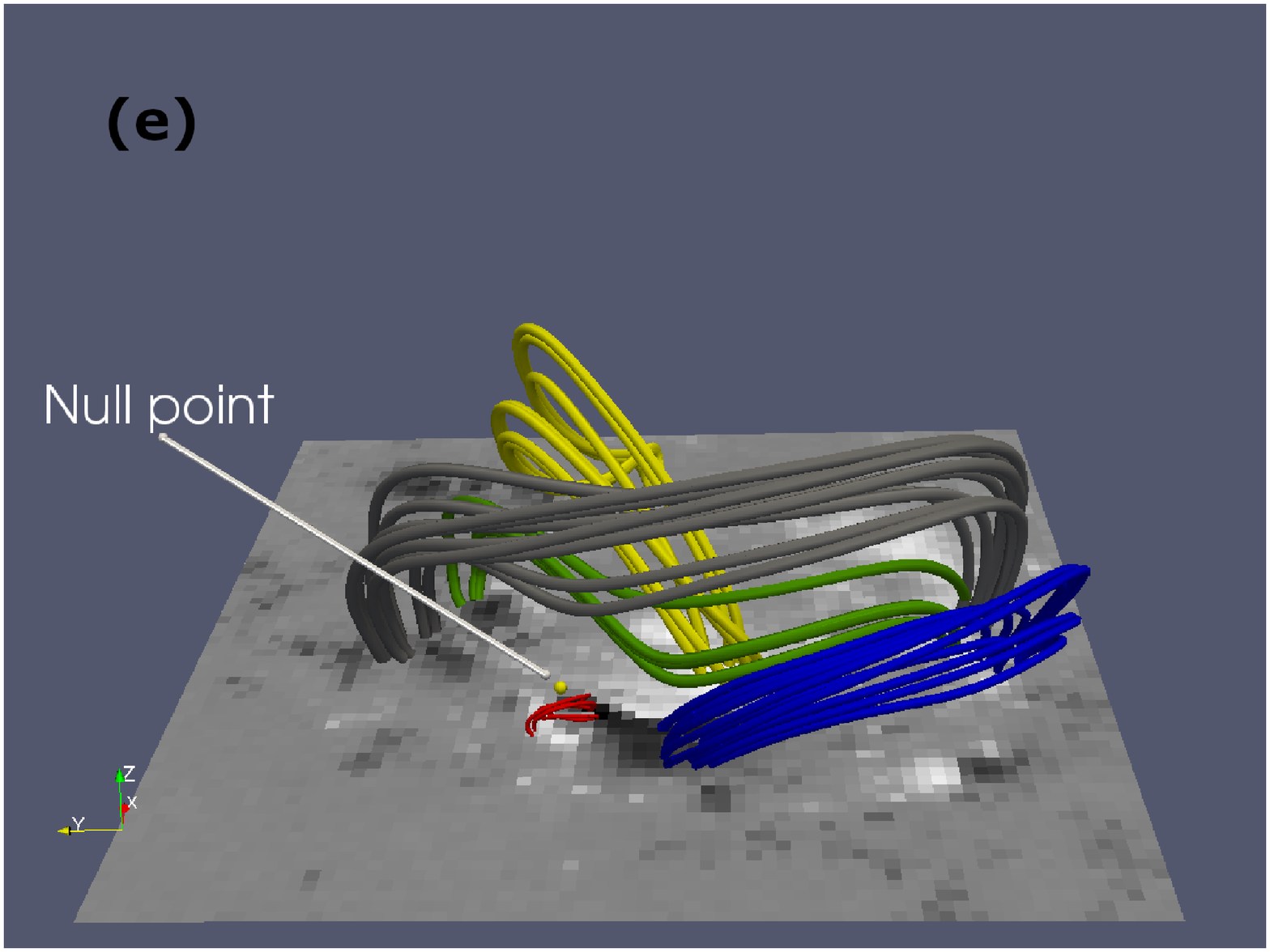}
    \includegraphics[trim=0 0 0 0, clip, scale=0.27]{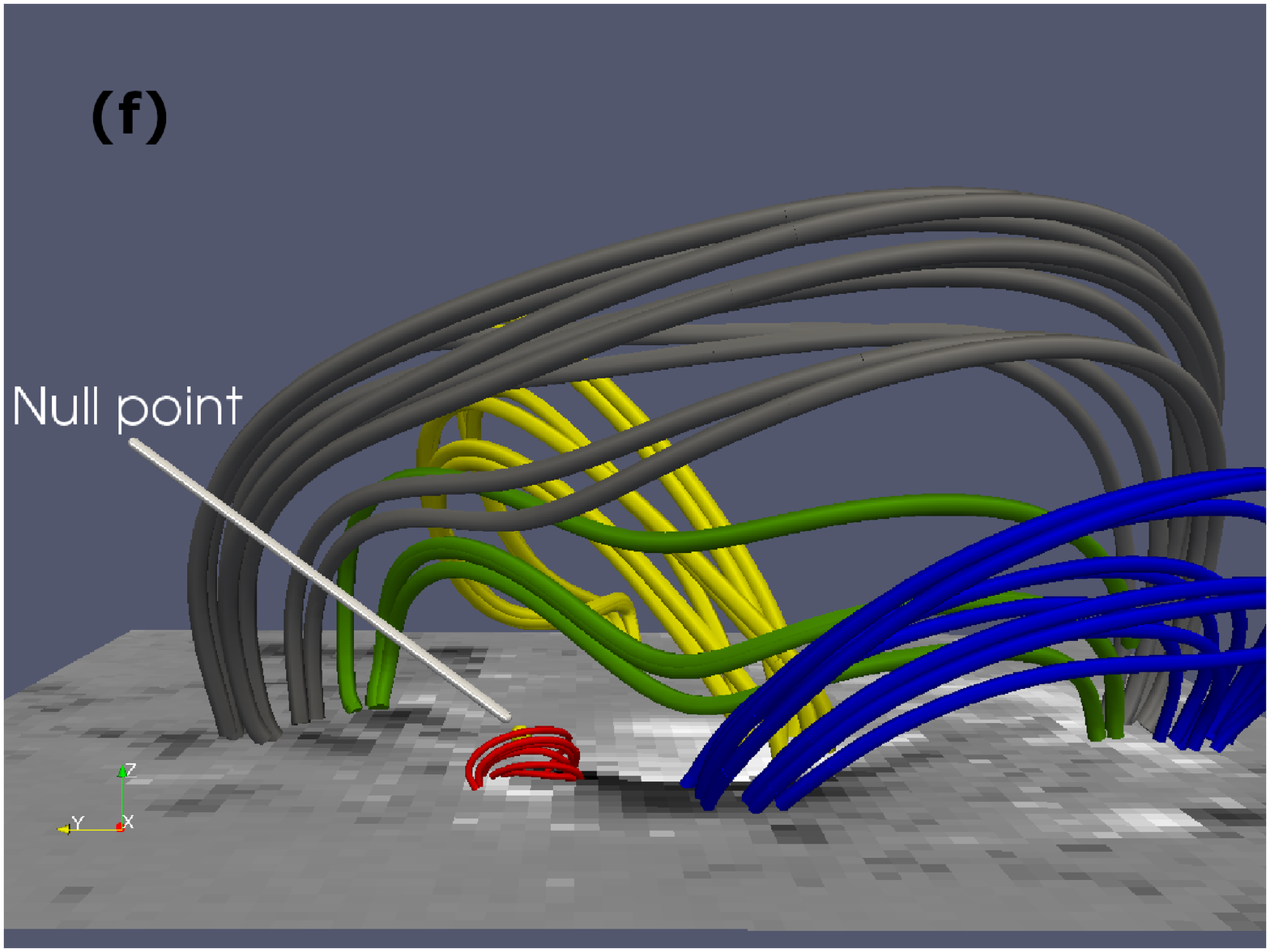}\\
  \caption{Non$-$linear force-free extrapolations of the AR NOAA 12673 obtained from vector magnetograms taken by SDO/HMI, as boundary conditions, on Sept. 06 at 00:00 UT (top raw), 08:48 UT (middle raw) and 11:48 UT (bottom raw). In the top panels we reported the indications of the main sunspots and umbrae already defined in Figure 1. In the top panel $``(a)"$ solar North is at the top and solar East is to the left. See text for more details.
   \label{Fig9}}
\end{figure*}

{\it Facilities:} SDO (HMI, AIA)


\end{article}

\end{document}